\documentclass[11pt,fleqn]{article}

\usepackage{a4}
\usepackage{amstext}
\usepackage{amsfonts}
\usepackage{amssymb}
\usepackage{color}
\usepackage{cite}
\usepackage{epsfig}
\usepackage{mathtools}
\usepackage{ulem}
\usepackage{subfigure}
\usepackage{framed}

\usepackage{authblk}

\newcommand{\Eqref}[1]{Eq.~\eqref{#1}}

\begin{document}


\begin{center}

{\huge \bf $\Lambda_{\overline{\textrm{MS}}}^{(n_f=2 )}$ from a momentum space analysis}

{\huge \bf of the quark-antiquark static potential}

\vspace{0.5cm}

\textbf{Felix Karbstein$^{1,2}$, Antje Peters$^3$, Marc Wagner$^{3,4}$}

$^1$~Helmholtz-Institut Jena, Fr\"obelstieg 3, D-07743 Jena, Germany

$^2$~Theoretisch-Physikalisches Institut, Friedrich-Schiller-Universit\"at Jena, Max-Wien-Platz 1, D-07743 Jena, Germany

$^3$~Goethe-Universit\"at Frankfurt am Main, Institut f\"ur Theoretische Physik, Max-von-Laue-Stra{\ss}e 1, D-60438 Frankfurt am Main, Germany

$^4$~European Twisted Mass Collaboration (ETMC)

\vspace{0.5cm}

August 25, 2014

\end{center}

\begin{tabular*}{16cm}{l@{\extracolsep{\fill}}r} \hline \end{tabular*}

\vspace{-0.4cm}
\begin{center} \textbf{Abstract} \end{center}
\vspace{-0.4cm}

We determine $\Lambda_{\overline{\textrm{MS}}}^{(n_f=2)}$ by fitting perturbative expressions for the quark-antiquark static potential to lattice results for QCD with $n_f=2$ dynamical quark flavors. To this end we use the perturbative static potential at the presently best known accuracy, i.e.\ up to ${\cal O}(\alpha_s^4)$, in momentum space. The lattice potential is computed on a fine lattice with $a \approx 0.042 \, \textrm{fm}$ in position space. To allow for a comparison and matching of both results, the lattice potential is transformed into momentum space by means of a discrete Fourier transform. The value of $\Lambda_{\overline{\textrm{MS}}}^{(n_f=2)}$ is extracted in momentum space. All sources of statistical and systematic errors are discussed.
The uncertainty in the value of $\Lambda_{\overline{\textrm{MS}}}^{(n_f=2)}$ is found to be smaller than that obtained in a recent position space analysis of the static potential based on the same lattice data.

\begin{tabular*}{16cm}{l@{\extracolsep{\fill}}r} \hline \end{tabular*}

\thispagestyle{empty}


\newpage

\setcounter{page}{1}

\section{\label{sec:intro} Introduction}

In this paper we aim at determining $\Lambda_{\overline{\rm MS}}$ by comparing lattice and perturbative results for the quark-antiquark ($Q\bar Q$) static potential%
\footnote{In agreement with the prevalent notation, particularly in the field of lattice QCD, we use the terms {\it static potential} and {\it static energy} synonymously.
Note, however, that in the literature sometimes a distinction is made and these terms refer to different quantities.}
in momentum space.
More precisely, we restrict ourselves to quantum chromodynamics (QCD) with $n_f=2$ dynamical quark flavors, i.e.\ exclusively focus on $\Lambda_{\overline{\textrm{MS}}}^{(n_f=2)}$.

The $Q\bar Q$ static potential $V$ amounts to the interaction energy of the color-singlet state made up of a static quark $Q$ and its antiquark $\bar Q$ separated by a distance $r=|\vec{r}|$ (in position space), or characterized by a momentum transfer $p=|\vec{p}|$ (in momentum space), respectively.
Accordingly, we denote the potential in position space by $V(r)$ and in momentum space by $V(p)$.

While many studies aiming at the extraction of $\Lambda_{\overline{\rm MS}}$ from the $Q\bar Q$ static potential have been performed so far, e.g.\ \cite{Michael:1992nj,Gockeler:2005rv,Brambilla:2010pp,Leder:2010kz,Jansen:2011vv,Bazavov:2012ka,Fritzsch:2012wq,Bazavov:2014soa}, to the best of our knowledge we are the first to transform the lattice data to momentum space and compare and match perturbative and lattice results in momentum space.
As will be discussed in detail below, our present study is mainly motivated by the significantly worse convergence behavior of the perturbative potential in position space as compared to momentum space \cite{Aglietti:1995tg,Jezabek:1998pj,Jezabek:1998wk,Beneke:1998rk}.

Apart from that, many other approaches aim at determining $\Lambda_{\overline{\rm MS}}$ or alternatively the strong coupling $\alpha_s$ at a specific momentum scale, e.g.\ the $Z$-mass scale.\footnote{Exclusively focusing on QCD with just two dynamical quark flavors, in the framework of the present study we favor the specification of $\Lambda_{\overline{\rm MS}}$ rather than $\alpha_s(M_Z)$.
A reasonable extraction of $\alpha_s(M_Z)$ from an $n_f=2$ study would at least require the discussion of flavor thresholds and extrapolations from $n_f\to n_f+1$, and in a sense obscure the main aim of our study.}
For recent lattice studies and results, see e.g.\ the lattice computations \cite{DellaMorte:2004bc,Shintani:2008ga,Aoki:2009tf,Sternbeck:2010xu,McNeile:2010ji,Blossier:2010ky,Sternbeck:2012qs,Blossier:2013ioa,Cichy:2013eoa} using the Schr\"odinger functional, vacuum polarization functions, ghost and gluon propagators, heavy quark correlators and the Dirac operator spectrum. Other works \cite{Abbate:2010xh,Kneur:2011vi,Boito:2012cr,Gehrmann:2012sc,Abbas:2012fi,Kneur:2013coa,Alekhin:2013nda,Andreev:2014wwa} focus e.g.\ on $\tau$ decays and electron-positron as well as electron-proton collisions.

Higher-order perturbative calculations in QCD are most conveniently performed in momentum space. This is particularly true for the perturbative $Q\bar Q$ static potential.
Hence, the perturbative expression for the static potential at the highest current accuracy, which encompasses all contributions up to ${\cal O}(\alpha_s^4)$, is directly accessible in momentum space.
Due to the fact that QCD is asymptotically free, perturbative calculations in QCD are viable only at large momentum transfers $p\gg\Lambda_{\rm QCD}$,
with $\Lambda_{\rm QCD}$ denoting the {\it QCD (momentum) scale}, which can be seen as separating the regimes of perturbative and non-perturbative physics.
However, note that in standard perturbation theory loop diagrams come along with integrations of the loop four-momentum over the full momentum regime, implying that such loops naturally also receive contributions from momenta $\lesssim\Lambda_{\rm QCD}$ for which perturbation theory is no longer trustworthy.
The leading uncontrolled contribution $\delta V(p)$ contained in the perturbative potential in momentum space arising from this kind of diagrams is quadratic in $\Lambda_{\rm QCD}$ and scales as $\delta V(p)\sim-\frac{4\pi\alpha_s}{p^2}\bigl(\frac{\Lambda_{\rm QCD}}{p}\bigr)^2$ \cite{Beneke:1998rk}.
Taking into account that $V(p)\sim-\frac{4\pi\alpha_s}{p^2}\bigl(1+{\cal O}(\alpha_s)\bigr)$ and $p\gg\Lambda_{\rm QCD}$, i.e.\ $\frac{\Lambda_{\rm QCD}}{p}\ll1$, this only amounts to a tiny correction.
For completeness, note that the strong coupling $\alpha_s$ depends on an {\it a priory} arbitrarily chosen renormalization momentum scale $\mu$, i.e.\ $\alpha_s=\alpha_s(\mu)$.
The $\mu$ dependence is such that $\alpha_s(\mu)\ll1$ only for $\mu\gg\Lambda_{\rm QCD}$. 
A particularly obvious choice of the renormalization scale for the couplings in $V(p)$ is $\mu=p$, corresponding to an identification of $\mu$ with the {\it typical momentum scale} of the quantity under consideration.

Conversely, within lattice QCD the $Q\bar Q$ static potential is naturally computed in position space.
It can be extracted straightforwardly by studying the exponential decay of the rectangular Wilson loop as a function of its temporal extension \cite{Brown:1979ya}.
Of course, lattice simulations at a given lattice spacing $a$ cannot resolve arbitrarily small separations.
Moreover, the minimum attainable lattice spacing is limited by the available computing power.
The behavior of the static potential at small $Q\bar Q$ separations is intimately related to its behavior at large momenta. Consequently, after a Fourier transform to momentum space, lattice results are expected to allow for reliable insights only below a certain threshold momentum,
which is $\lesssim$ the maximum momentum $\overline{p}=\pi/a$ that can be resolved on a lattice of spacing $a$ (cf. Sec.~\ref{SEC572} below).

Taking into account the above constraints, a comparison and matching of perturbative and lattice results in momentum space is limited to momenta $p$ fulfilling $\Lambda_{\rm QCD}\ll p\ll \overline{p}$.

Analogous considerations can be invoked to delimit the fitting interval in position space (cf.\ e.g.\ \cite{Jansen:2011vv}),
employing that the typical momentum scale that can be attributed to a relative distance $r$ scales as $p\sim1/r$. In position space the manifestly perturbative regime is thus characterized by  $1/r\gg\Lambda_{\rm QCD}$.
However, an important difference is that the leading uncontrolled contribution $\delta V(r)$ to the static potential in position space as determined by a standard Fourier transform from momentum to position space is more pronounced than in momentum space: it is linear in $\Lambda_{\rm QCD}$, scales as $\delta V(r)\sim\frac{\alpha_s}{r}(r\Lambda_{\rm QCD})=\alpha_s\Lambda_{\rm QCD}$,
and arises exclusively from the low momentum part of the Fourier integral over momenta $\lesssim\Lambda_{\rm QCD}$, for which perturbation theory is no longer trustworthy \cite{Beneke:1998rk}%
\footnote{Recall that in the perturbative regime both dimensionless quantities, $\Lambda_{\rm QCD}/p$ and $r\Lambda_{\rm QCD}$, are small and of the same order of magnitude, i.e.\ $\Lambda_{\rm QCD}/p\sim r\Lambda_{\rm QCD}\sim\epsilon\ll1$.
Correspondingly, $r\Lambda_{\rm QCD}={\cal O}(\epsilon)$, while $\bigl(\frac{\Lambda_{\rm QCD}}{p}\bigr)^2={\cal O}(\epsilon^2)$.}.
For the position space potential a seemingly obvious choice of the renormalization scale is $\mu=1/r$, which amounts to the {\it typical momentum scale} to be associated with the static quarks separated by a distance $r$.

However, it has been recognized long ago that in particular for the identification $\mu=1/r$ the convergence of the perturbative potential in position space as defined by a standard Fourier transform from momentum space is spoiled \cite{Billoire:1979ih},
while it can be significantly improved by subtracting just the uncontrolled contribution $\delta V(r)$ linear in $\Lambda_{\rm QCD}$ \cite{Beneke:1998rk}.
Unfortunately this necessitates the specification of an additional {\it subtraction scale} and thereby increases the number of free parameters.
Without any subtraction procedure a meaningful fit of perturbative expressions for the $Q\bar Q$ static potential to lattice results in position space is not possible:
the perturbative expressions only show a controlled convergence behavior for separations, which are much smaller than the minimum accessible separations on state-of-the-art lattice simulations
(cf.\ e.g.\ \cite{Pineda:2001zq,Pineda:2002se,Laschka:2011zr,Karbstein:2013zxa}). 
Correspondingly, aiming at an accurate determination of $\Lambda_{\overline{\rm MS}}$ by comparing the results from lattice QCD simulations with perturbative calculations of the $Q\bar Q$ static potential in position space,
considerable efforts are needed to cope with this issue. Various strategies to retain or restore the significantly better convergence of $V(p)$ also for $V(r)$ have been devised in the literature \cite{Beneke:1998rk,Pineda:2001zq,Pineda:2002se,Brambilla:2009bi,Brambilla:2010pp,Laschka:2011zr,Karbstein:2013zxa}.

An extraction of $\Lambda_{\overline{\rm MS}}$ directly in momentum space of course does not involve a Fourier transform of the perturbative potential, such that one may hope to circumvent most of these limitations. 
On the other hand, one now has to Fourier transform the lattice potential, which might eventually lead to similar problems.
As we will argue and demonstrate in detail in this paper, most favorable for us the latter concerns are not substantiated.

Our paper is organized as follows. Section~\ref{SEC572} is devoted to the $Q\bar Q$ static potential on the lattice. After briefly reviewing its computation in position space, we point out how we transform it to momentum space.
In Sec.~\ref{sec:perturbation} we summarize the present knowledge of the perturbative $Q\bar Q$ static potential and detail on the role of $\Lambda_{\overline{\rm MS}}$.
Special emphasis is put on the convergence behavior of both the perturbative $Q\bar Q$ static potential for $n_f=2$ in momentum space and the QCD $\beta$-function.
Section~\ref{SEC600} constitutes the main section of our paper. Here we describe in detail our momentum space analysis to extract $\Lambda_{\overline{\textrm{MS}}}^{(n_f=2 )}$ by fitting the perturbative expressions for the $Q\bar Q$ static potential $V(p)$ to the corresponding lattice results. The various error sources are identified and delineated and our final result for $\Lambda_{\overline{\textrm{MS}}}^{(n_f=2 )}$ is specified. Moreover, comparisons with the result of \cite{Jansen:2011vv}, amounting to a position space extraction of $\Lambda_{\overline{\textrm{MS}}}^{(n_f=2 )}$ based on the same lattice data, are made. Finally, we end with conclusions in Sec.~\ref{sec:Concl}.


\newpage
 
\section{\label{SEC572}The $Q\bar Q$ static potential in momentum space from lattice QCD }


\subsection{\label{SEC852}Gauge link configurations}

We use the same $n_f = 2$ gauge link configurations as for a recent determination of $\Lambda_{\overline{\textrm{MS}}}$, where, in contrast to this work, the lattice results and perturbative expressions for the static potential were compared and matched in position space \cite{Jansen:2011vv}. These gauge link configurations were generated by the European Twisted Mass Collaboration (ETMC) \cite{Boucaud:2007uk,Boucaud:2008xu,Baron:2009wt} with the tree-level Symanzik improved gauge action \cite{Weisz:1982zw},
\begin{equation}
S_\mathrm{G}[U] = \frac{\beta}{6} \bigg(b_0 \sum_{x,\mu\neq\nu} \textrm{Tr}\Big(1 - P^{1 \times 1}(x;\mu,\nu)\Big) + b_1 \sum_{x,\mu\neq\nu} \textrm{Tr}\Big(1 - P^{1 \times 2}(x;\mu,\nu)\Big)\bigg)
\end{equation}
with $b_0 = 1 - 8 b_1$ and $b_1 = -1/12$ and the Wilson twisted mass quark action \cite{Frezzotti:2000nk,Frezzotti:2003ni,Frezzotti:2004wz,Shindler:2007vp},
\begin{equation}
\label{EQN963} S_\mathrm{F}[\chi,\bar{\chi},U] = a^4 \sum_x \bar{\chi}(x) \Big(D_{\rm W} + i \mu_\mathrm{q} \gamma_5 \tau_3\Big) \chi(x)
\end{equation}
with
\begin{equation}
D_\mathrm{W} = \frac{1}{2} \Big(\gamma_\mu \Big(\nabla_\mu + \nabla^\ast_\mu\Big) - a \nabla^\ast_\mu \nabla_\mu\Big) + m_0 .
\end{equation}
Here $a$ denotes the lattice spacing, $\nabla_\mu$ and $\nabla^\ast_\mu$ are the gauge covariant forward and backward derivatives, $m_0$ and $\mu_\mathrm{q}$ are the bare untwisted and twisted quark masses, $\tau_3$ is the third Pauli matrix acting in flavor space, and $\chi = (\chi^{(u)} , \chi^{(d)})$ represents the quark fields in the so-called twisted basis.

The twist angle $\omega$ is given by $\omega = \arctan(\mu_\mathrm{R} / m_\mathrm{R})$, where $\mu_\mathrm{R}$ and $m_\mathrm{R}$ denote the renormalized twisted and untwisted quark masses. 
For the ensembles of gauge link configurations considered in the present study (cf.~Table~\ref{TAB077}) $\omega$ has been tuned to $\pi / 2$ by adjusting $m_0$ appropriately.
This ensures automatic $\mathcal{O}(a)$ improvement for many observables including the static potential (cf.\ \cite{Boucaud:2008xu} for details).

The considered gauge link configurations cover several different values of the lattice spacing, the pion mass $m_\textrm{PS}$ and the spacetime volume $L^3 \times T$; cf.~Table~\ref{TAB077}, which also provides the number of gauge link configurations, used for the computation of the static potential, for each ensemble. The lattice spacing in physical units has been set via the pion mass and the pion decay constant, using chiral perturbation theory. The resulting value for the
hadronic scale\footnote{The hadronic scale $r_0$ is defined via $r_0^2 F(r_0) = 1.65$, with $F(r) = {\rm d}V(r) / {\rm d}r$ \cite{Sommer:1993ce}.} $r_0$
is $r_0 = 0.420(14) \, \textrm{fm}$ (cf.\ Sec.~5 of \cite{Boucaud:2008xu} and Tab.~8 of \cite{Baron:2009wt}). For further details on the generation of these gauge field configurations as well as on the computation and the analysis of standard quantities (e.g.\ lattice spacing and pion mass) we refer the reader to \cite{Boucaud:2008xu,Baron:2009wt}.




 










\begin{table}[htb]
\begin{center}
\begin{tabular}{|c|c|c|c|c|c|}
\hline
 & & & & & \vspace{-0.40cm} \\
$\beta$ & $a$ in $\textrm{fm}$ & $(L/a)^3 \times T/a$ & $m_\textrm{PS}$ in $\textrm{MeV}$ & $r_0 / a$ & \# gauges \\
 & & & & & \vspace{-0.40cm} \\
\hline
 & & & & & \vspace{-0.40cm} \\
\hline
 & & & & & \vspace{-0.40cm} \\
$3.90$ & $0.079(3)\phantom{00}$ & $24^3 \times 48$ & $340(13)$ & $5.36(4)\phantom{0}$ & $168$ \\
 & & & & & \vspace{-0.40cm} \\
\hline
 & & & & & \vspace{-0.40cm} \\
$4.05$ & $0.063(2)\phantom{00}$ & $32^3 \times 64$ & $325(10)$ & $6.73(5)\phantom{0}$ & $\phantom{0}71$ \\
%
%
 & & & & & \vspace{-0.40cm} \\
\hline
 & & & & & \vspace{-0.40cm} \\
$4.20$ & $0.0514(8)\phantom{0}$ & $48^3 \times 96$ & $284(5)\phantom{0}$ & $8.36(6)\phantom{0}$ & $\phantom{0}46$ \\
 & & & & & \vspace{-0.40cm} \\
\hline
 & & & & & \vspace{-0.40cm} \\
$4.35$ & $0.0420(17)$           & $32^3 \times 64$ & $352(22)$ & $9.81(13)$ & $146$\vspace{-0.40cm} \\
 & & & & & \\
\hline
\end{tabular}

\caption{\label{TAB077}Ensembles of gauge link configurations employed in the present study.}

\end{center}
\end{table}


\subsection{\label{EQN808}Computation of the $Q\bar Q$ static potential in position space}

First we determine the $Q\bar{Q}$ static potential $V(\vec{r})$ in position space. In a second step, $V(\vec{r})$ is transformed to momentum space by means of a discrete Fourier transform.

To be able to perform this Fourier transform numerically we need the static potential for all $\vec{r} = \vec{n} a$ inside a finite periodic spatial volume $L'^3$ of side length $L'$. This is achieved as follows: First, we compute $V(\vec{n} a)$ with $n_i\in\{0,1,\ldots,N'/2\}$, where $N'$ is even and defined as $N' \equiv L'/a$. Second, we realize the periodicity by defining $V(n_x a , n_y a , n_z a) \equiv V(|n_x| a , |n_y| a , |n_z| a)$, where now $n_i \in\{-\frac{N'}{2} + 1 , -\frac{N'}{2} + 2 , \ldots ,\frac{N'}{2}\}$.

To keep finite volume effects on a negligible level, the spatial volume $L'^3$ needs to be sufficiently large (cf. also point (2) below).
While a lattice computation of the static potential $V(\vec{n} a)$ for all $n_x , n_y , n_z = 0 , 1 , \ldots , N'/2$ is possible in principle, it is extremely computer time consuming in practice: One needs to generate gauge link configurations for such large volumes and has to compute both on- and off-axis Wilson loops for all possible quark-antiquark separations $\vec{r}=\vec{n} a$. On the other hand the shape of the static potential at large separations is known to be accurately described by $V(r) = A_0 + \sigma r + A_1/r$, where $A_0$ denotes a constant offset parameter, $\sigma$ is the string tension and $A_1 \gtrsim -\pi/12$ \cite{Luscher:1980fr,Luscher:1980ac,Donnellan:2010mx}.

Several clarifying comments are in order here.
Accounting for a nonvanishing number $n_f$ of light dynamical quark flavors,
there exists a certain threshold distance $r_c\gtrsim1\,{\rm fm}$ such that for separations $r>r_c$ of the static quark $Q$ and its antiquark $\bar Q$ the $Q\bar Q$ state is energetically disfavored in comparison to a pair of static-light mesons, $B\bar B$ \cite{Bali:2005fu}. This effect is known as string breaking.
The ground state of the system is made up of dynamical quarks, gluons and static quarks $Q$ and $\bar Q$ separated by a distance $r$. 
It scales string-like  $\sim \sigma r$ in the nonperturbative regime below $r_c$, but becomes completely independent of $r$ for $r>r_c$, and saturates at about twice the $B$ meson mass.
However, the $Q\bar Q$ state scaling $\sim \sigma r$ can still be traced for $r>r_c$ also, where it corresponds to an excited state:
Apart from resulting in a narrow mini-gap feature at $r\approx r_c$, i.e.\ where the energy of the $Q\bar Q$ state equals that of the $B\bar B$ state, mixing effects between the $Q$ and $B$ sectors are tiny.
Wilson loops as also employed here to extract the static potential on the lattice [cf. \Eqref{EQN600} below] are particularly insensitive to the $B$ sector.
On the other hand, standard perturbation theory for the $Q\bar Q$ static potential manifestly focuses on the $Q$ sector of the theory:
While it accounts for dynamical light quarks in loop diagrams, in this framework the virtual light quarks can never become real.
For these reasons, even though there occurs string breaking for $n_f\neq0$, when focusing on the $Q\bar Q$ potential $V(r)$ 
we manifestly limit ourselves to the $Q$ sector, i.e.\ focus on the $r$ dependent, linearly rising component of the potential also for $r>r_c$.

\begin{figure}[htb]
\begin{center}
\includegraphics[width=0.8\textwidth]{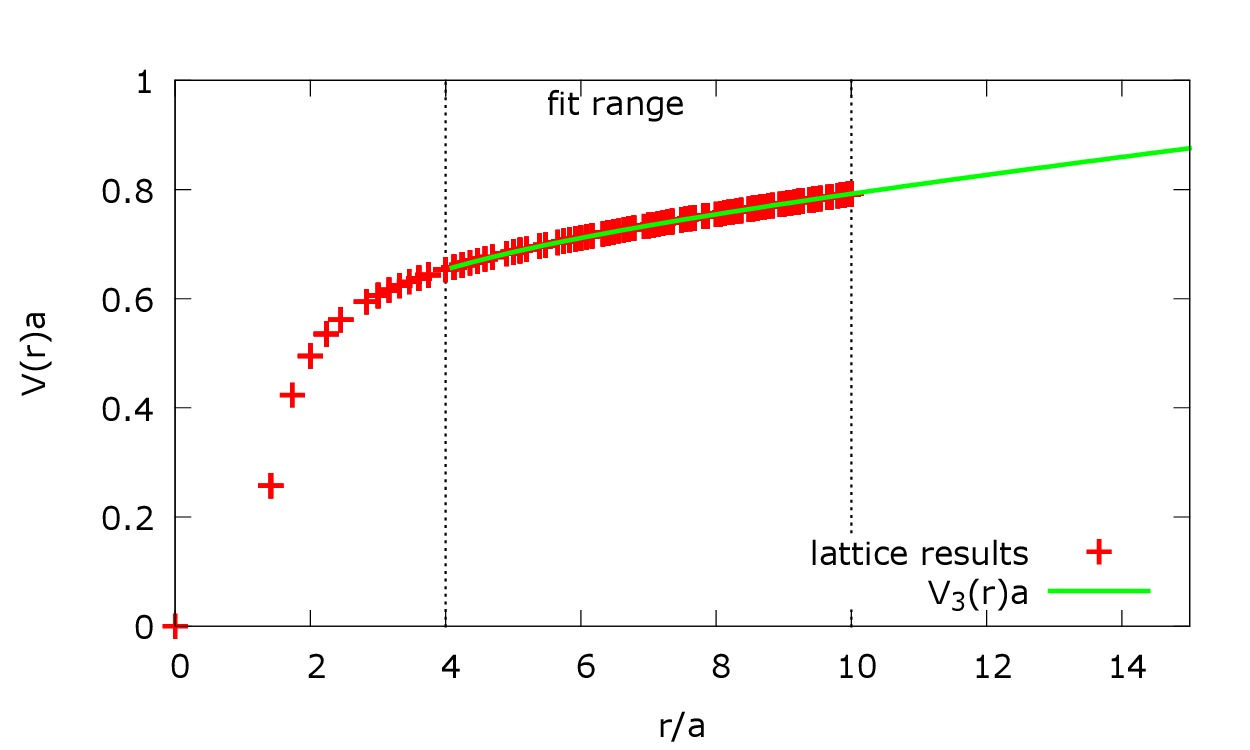}
\caption{\label{FIG600}Position space potential $V_3(r)$ (green curve) obtained by a $\chi^2$ minimizing fit to the lattice results $V(\vec{r})$ (red dots) in the region $4 a \leq |\vec{r}| \leq 10 a$ with $a \approx 0.0420 \, \textrm{fm}$. Note that the statistical errors (actually also depicted here) are so tiny that the curve for $V_3(r)$ and its error band fall on top of each other and are indiscernible by eye.}
\end{center}
\end{figure}

Moreover, the large distance behavior of $V(r)$ is expected to have a rather weak effect on the Fourier transformed potential for $p\gg\Lambda_{\rm QCD}$, i.e.\ the momentum 
regime used for the matching to perturbation theory and the $\Lambda_{\overline{\textrm{MS}}}$ determination in this work.\footnote{To substantiate this rather vague statement given here, we have explicitly checked and confirmed that the large distance behavior of the lattice potential $V(r)$ has only a very mild influence of the value of $\Lambda_{\overline{\rm MS}}$ to be extracted from the $Q\bar Q$ static potential in momentum space, by modeling the long distance behavior of $V(r)$ with different functional forms (cf. point (2) below, and the numerical results in Sec.~\ref{input_var}, particularly Table~\ref{TAB015}).} Therefore, we stick to the following strategy:
\begin{itemize}
\item[(1)] \textbf{Perform a standard lattice computation of }$V(\vec{r})$\textbf{ for }$|\vec{r}|=|\vec{n}|a \leq r_\textrm{max}$\textbf{:} \\
For quark-antiquark separations $|\vec{n}| a \leq r_\textrm{max}$ we extract $V(\vec{r})$ from the exponential decay of Wilson loop averages $\langle W(\vec{r},t) \rangle$ with respect to their temporal extent $t$, while keeping their spatial extent $r$ fixed.
To this end we first compute
\begin{equation}
 V^\textrm{(effective)}(\vec{r},t) =  \frac{1}{a} \ln\bigg(\frac{\langle W(\vec{r},t) \rangle}{\langle W(\vec{r},t+a) \rangle}\bigg) .
\end{equation}
Somewhat arbitrarily, we choose $r_\textrm{max}\approx0.42\,{\rm fm}$ corresponding to $r_\textrm{max} = 10a$ for our smallest lattice spacing (cf. Table~\ref{TAB077}).
In a second step the $t$-independent quantity $V(\vec{r})$ is obtained by performing an uncorrelated $\chi^2$ minimizing fit to $V^\textrm{(effective)}(\vec{r},t)$ in a suitable $t$ range. This range is chosen such that excited states are strongly suppressed, while statistical errors are still small.

We use the ensembles listed in Table~\ref{TAB077} and consider on- and off-axis Wilson loops formed by APE smeared spatial links ($N_\textrm{APE} = 60$, $\alpha_\textrm{APE} = 0.5$ for all our gauge link ensembles) and ordinary, i.e .\ unsmeared temporal links. For a detailed explanation regarding the construction of off-axis Wilson loops cf.\ \cite{Jansen:2011vv}. For a definition of APE smearing we refer to \cite{Jansen:2008si}.

\item[(2)] \textbf{Model }$V(\vec{r})$\textbf{ for }$|\vec{r}|=|\vec{n}| a > r_\textrm{max}$\textbf{ and }$|r_i| = |n_i|a \leq L'/2$\textbf{:} \\
For quark-antiquark separations $|\vec{n}| a > r_\textrm{max}$ while $|n_i|a \leq L'/2$ we model the lattice potential by
\begin{equation}
\label{EQN600} V(\vec{r}) = V_M(r)  \equiv  A_0 + \sigma r + \sum_{m=1}^M \frac{A_m}{r^m} .
\end{equation}
For our finest lattice spacing $a \approx 0.0420 \, \textrm{fm}$ we use $L'/a = 256$, obviously fulfilling $L' \gg L$ (cf. Table~\ref{TAB077}).
To ensure that the extracted value of $\Lambda_{\overline{\rm MS}}$ is independent of the choice for $L'$, we also performed computations with $L'/a = 128$ and $L'/a = 512$ and found essentially identical results for $\Lambda_{\overline{\rm MS}}$
(the deviations are below $1\,{\rm MeV}$).

In Sec.~\ref{SEC600} different values of $M\in\{1,2,3,4\}$ are used to quantify systematic errors associated with this modeling of the long range part of the lattice potential. The string tension is fixed to $\sigma = 1550 \, \textrm{MeV}/\textrm{fm}$ (corresponding to $r_0 = 0.420 \, \textrm{fm}$ determined on the same gauge link configurations we are using throughout this work \cite{Baron:2009wt}). While $A_1 = -\pi/12$ in the bosonic string picture \cite{Luscher:1980fr,Luscher:1980ac}, lattice simulations with $n_f=2$ quark flavors yield a larger value $A_1 \approx -0.3 \ldots -0.5$ \cite{Donnellan:2010mx}. We determine $A_m$ with $m\in\{0,1,\ldots,M\}$ by a $\chi^2$ minimizing fit of \Eqref{EQN600} to the lattice results determined in step (1) in the region $r_\textrm{min} \leq r \leq r_\textrm{max}$. In order to have $\chi^2\lesssim1$, for our smallest lattice spacing $a = 0.0420 \, \textrm{fm}$ we choose $r_{\rm min}=8a$ for $M=1$,  $r_{\rm min}=6a$ for $M=2$, and $r_\textrm{min} =4 a$ for $M =\{3,4\}$. 

The resulting function $V_3(r)$ for $M=3$ is shown in Figure~\ref{FIG600} together with the lattice results for $V(\vec{r})$.
\end{itemize}

\subsection{Computation of the $Q\bar Q$ static potential in momentum space}

We define the $Q\bar Q$ static potential in momentum space, $V(\vec{p})$ with $\vec{p} = 2 \pi \vec{k} / L'$, by the discrete Fourier transform of $V(\vec{r})$,
\begin{equation}
\label{EQN601} V(\vec{p}) = V(2 \pi \vec{k} / L') = \sum_{n_x , n_y , n_z} a^3 \exp\bigg(-\frac{2 \pi i \vec{k} \vec{n}}{N'}\bigg) V(\vec{n} a) ,
\end{equation}
where the sum is also over all possible values of $n_i\equiv-\frac{N'}{2}+\nu_i$ with $\nu_i\in\{1,2,\ldots,N'\}$, and $k_i\equiv -\frac{N'}{2} + \kappa_i$ with $\kappa_i\in\{1,2,\ldots,N'\}$.

In the continuum and in infinite volume $V(\vec{p})$ is rotationally symmetric, i.e.\ $V(\vec{p})=V(p)$.
Since both a cubic lattice discretization and a cubic periodic volume break rotational symmetry, this is no longer true for the lattice potential~\eqref{EQN601}.
The deviations from the rotationally invariant continuum and infinite volume potential are expected to be particularly small, when restricting the dimensionless lattice momenta $\vec k$ to values inside a cylinder of unit radius around the lattice diagonal, i.e.\ demanding
\begin{equation}
\vec{k}^2 - (\vec{k} \vec{d})^2  \leq 1\, , \quad \vec{d} = \frac{1}{\sqrt{3}} (1 , 1 , 1)\, .
\end{equation}
This so-called cylinder cut is frequently used in lattice momentum space computations of propagators. For details cf.\ \cite{Sternbeck:2006rd}.

The maximal momentum value along each of the three axes is $\overline{p} = \pi / a \approx 15 \, \textrm{GeV}$ for our smallest lattice spacing $a \approx 0.0420 \, \textrm{fm}$. Similar to position space, where the minimum on-axis separation is $a$ and the static potential is essentially free of discretization errors for $r \gtrsim 3 a$, one might expect rather small discretization errors for $p \lesssim \overline{p} / 3 \approx 5 \, \textrm{GeV}$. These expectations will be confirmed in Sec.~\ref{input_var} below, where in particular the results depicted in Figure~\ref{FIG003} (c) and (d) show that a variation of the maximum momentum employed in the extraction of $\Lambda_{\overline{\rm MS}}$ in a range $\lesssim3 \, \textrm{GeV}$
basically does not change the value of $\Lambda_{\overline{\rm MS}}$.

\begin{figure}[htb]
\begin{center}
\includegraphics[width=0.8\textwidth]{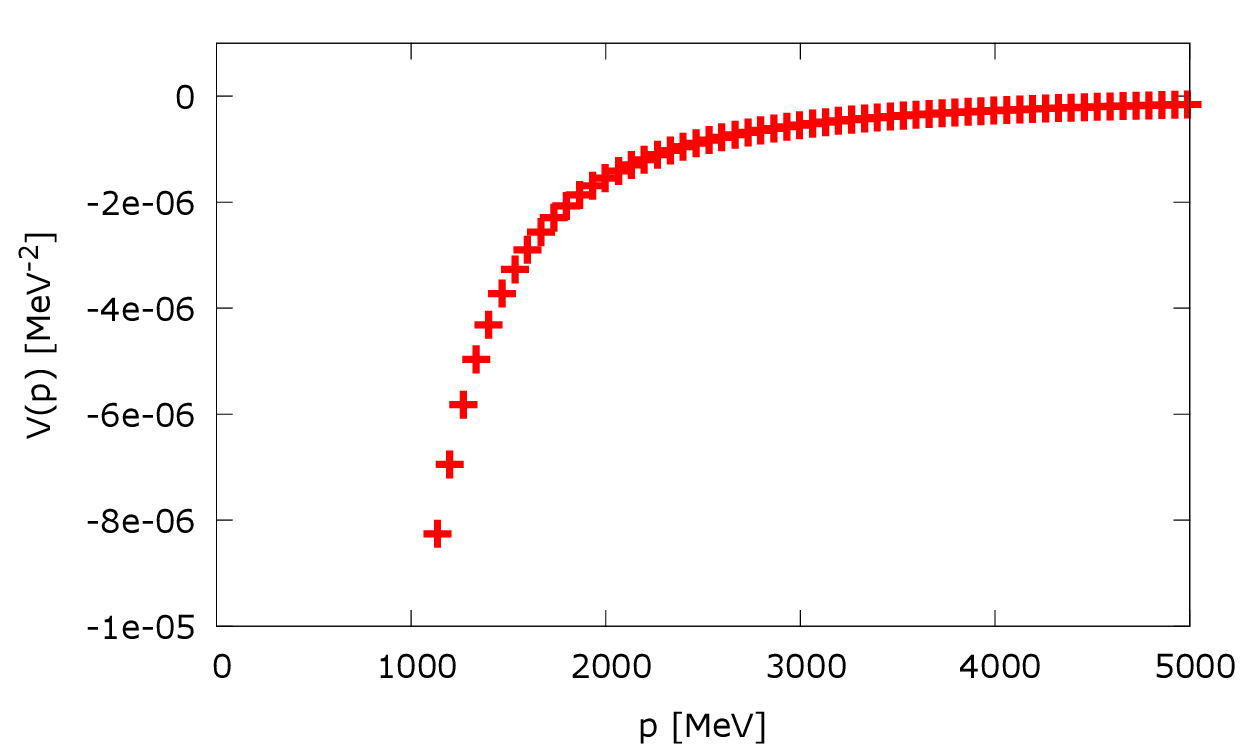}
\caption{\label{FIG601}$V(p)$ for $a \approx 0.0420 \, \textrm{fm}$ obtained from $V_3(r)$ after applying the cylinder cut.}
\end{center}
\end{figure}

The final result $V(p)$ for our smallest lattice spacing $a \approx 0.0420 \, \textrm{fm}$, obtained with $V_3(r)$ [cf. \Eqref{EQN600}  and Figure~\ref{FIG600}] is shown in Figure~\ref{FIG601}\footnote{Throughout this paper all computations are performed in units of the lattice spacing $a$, e.g.\ we work with quantities $aV(r)$, $V(p) / a^2$ and $a\Lambda_{\overline{\rm MS}}$, which are independent of any potential errors or uncertainties regarding scale setting; for a recent review cf.\ \cite{Sommer:2014mea}. Nevertheless, the axes in the presented plots as well as the numbers quoted in the main text often given in units of $\textrm{MeV}$ or $\textrm{fm}$. This is intended to make the physical scales more obvious. To this end we use $a = 0.0420 \, \textrm{fm}$ for our smallest lattice spacing corresponding to $\beta = 4.35$ (cf.\ also Table~\ref{TAB077}).}.

\newpage

\section{\label{sec:perturbation}Perturbative insights into the $Q\bar Q$ static potential}

\subsection{The $Q\bar Q$ static potential in perturbation theory}

To allow for easier reference and to keep this paper self-contained, we briefly summarize the present status of the $Q\bar Q$ static potential in perturbation theory. 
Quantities which depend on the particular renormalization scheme used are given in the $\overline{\rm MS}$ scheme \cite{Bardeen:1978yd,Furmanski:1981cw}.

In the perturbative momentum regime, i.e.\ for momenta $p=|\vec{p}|\gg\Lambda_{\rm QCD}$, the $Q\bar Q$ static potential is conveniently represented as 
\begin{equation}
 V(p)=-C_F\frac{4\pi}{p^2}\alpha_V[\alpha_s(\mu),L(\mu,p)] \label{eq:V(p)}
\end{equation}
with $C_F$ the eigenvalue of the quadratic Casimir operator for the fundamental representation of the gauge group; $C_F=4/3$ for ${\rm SU}(3)$.
Pulling out the overall factor $\sim1/p^2$ the entire non-trivial structure of $V(p)$ can be encoded in the dimensionless quantity $\alpha_V$,
which in turn is a function of both the coupling $\alpha_s(\mu)$ and $L\equiv L(\mu,p)=\ln\frac{\mu^2}{p^2}$. $\alpha_s(\mu)$ is evaluated at an {\it a priori} arbitrarily chosen renormalization scale $\mu$ in the perturbative regime, i.e.\ $\mu\gg\Lambda_{\rm QCD}$, guaranteeing that $\alpha_s(\mu)\ll1$.

The running of the coupling $\alpha_s(\mu)$ as a function of the renormalization scale $\mu$ is governed by the QCD $\beta$-function defined as
\begin{equation}
 \beta[\alpha_s(\mu)]\equiv\frac{\mu}{\alpha_s(\mu)}\frac{d}{d \mu}\alpha_s(\mu)\,, \label{eq:beta}
\end{equation}
whose series expansion in powers of $\alpha_s$ is presently known with the following accuracy,
\begin{equation}
 \beta(\alpha_s)=-\frac{\alpha_s}{2\pi}\beta_0\biggl[1+\frac{\alpha_s}{4\pi}\frac{\beta_1}{\beta_0}+\biggl(\frac{\alpha_s}{4\pi}\biggr)^2\frac{\beta_2}{\beta_0}+\biggl(\frac{\alpha_s}{4\pi}\biggr)^3\frac{\beta_3}{\beta_0}
+ \ldots\,\biggr]\,. \label{eq:betaseries}
\end{equation}
While the expansion coefficients $\beta_0$ and $\beta_1$ are independent of the renormalization scheme,
$\beta_2$ and $\beta_3$ are scheme-dependent. They have been
determined for arbitrary compact semi-simple Lie groups in the $\overline{\rm
MS}$ scheme \cite{vanRitbergen:1997va}. 
For ${\rm SU}(3)$ with $n_f=2$ massless dynamical quark flavors they read
\begin{equation}
 \beta_0 =\frac{29}{3}\,, \quad \beta_1= \frac{230}{3}\,,\quad \beta_2 = \frac{48241}{54}\,,\quad \beta_3 = \frac{18799309}{1458}+\frac{275524}{81}\zeta(3) .\!
\end{equation}
The same quantities for arbitrary values of $n_f$ can be found e.g.\ in \cite{Jansen:2011vv}.

As the static potential is a physical observable, it should of course be independent of the explicit value of the renormalization scale $\mu$
and form a renormalization group (RG) invariant, i.e.\ fulfill
\begin{equation}
 \mu\frac{d}{d\mu}V(p)=0\,. \label{eq:RG_V}
\end{equation}
One might wonder, how this can come about with $\alpha_V$ in \Eqref{eq:V(p)} being a function of the two $\mu$-dependent quantities $\alpha_s(\mu)$ and $L(\mu,p)$.
However, knowing $V(p)$ -- and thus $\alpha_V$ -- at a certain 
accuracy in perturbation theory, e.g.\ up to ${\cal O}(\alpha_s^{\bar k})$,
\Eqref{eq:RG_V} only has to hold to this order, i.e.\
\begin{equation}
 \mu\frac{d}{d\mu}\alpha_V[\alpha_s,L]={\cal O}(\alpha_s^{\bar k+1})
 \quad\quad\leftrightarrow\quad\quad\left(\frac{\partial
} { \partial L}+\frac{\alpha_s}{2}\beta(\alpha_s)\frac{\partial}{\partial
\alpha_s}\right)\alpha_V[\alpha_s,L]={\cal O}(\alpha_s^{\bar k+1}) \,. \label{eq:RG_alphaV}
\end{equation}
Presently, all terms are known explicitly for $\bar k=4$, and $\alpha_V$ is of the following form,
\begin{multline}
 \alpha_{V}[\alpha_s(\mu),L(\mu,p)] = \alpha_s(\mu)\left\{1+\frac{\alpha_s(\mu)}{4\pi}\,P_1(L)
+\left(\frac{\alpha_s(\mu)}{4\pi}\right)^2P_2(L)\right. \\
\left.+\left(\frac{\alpha_s(\mu)}{4\pi}
\right)^3\Bigl[P_3(L)+a_{3 {\rm
ln}}\ln{\alpha_s}(\mu)\Bigr]+\ldots
\right\} . \label{eq:pert1}
\end{multline}
While $\alpha_V$ has a strict power-series expansion in $\alpha_s$ up to ${\cal O}(\alpha_s^3)$, beyond this order one also encounters logarithmic contributions in $\alpha_s$ \cite{Appelquist:1977es}, the first such term being $\sim \alpha_s^4\ln{\alpha_s}$ \cite{Brambilla:1999qa}.

Equation~\eqref{eq:RG_alphaV} constrains the $P_k(L)$ accounting for the entire $L$ dependence of $\alpha_V$ in \Eqref{eq:pert1} to be polynomials in $L$ of degree $k$, i.e.
\begin{equation}
 P_k(L)=\sum_{m=0}^{k}\rho_{km}L^m \label{eq:Pk}
\end{equation}
with dimensionless expansion coefficients $\rho_{km}$, and implies that, apart from the explicit values of $a_k\equiv\rho_{k0}=P_k(0)$ with $a_0=1$, the $\rho_{km}$ are fully determined by the coefficients of the $\beta$-function \cite{Chishtie:2001mf,Karbstein:2013zxa}. Their explicit expressions for $k\leq3$ are given in our notations in \cite{Karbstein:2013zxa}.

The coefficients $a_1$ \cite{Fischler:1977yf,Billoire:1979ih} and $a_2$ \cite{Peter:1996ig,Peter:1997me,Schroder:1998vy} are known analytically.
For gauge group ${\rm SU}(3)$, $n_f=2$ and in the $\overline{\rm MS}$-scheme they read 
\begin{equation}
 a_1=\frac{73}{9}\,, \quad a_2=\frac{25139}{162}+9\pi^2\Bigl(4-\frac{\pi^2}{4}\Bigr)+\frac{94}{3}\zeta(3) \,.
\end{equation}
Also the coefficients $a_3$ and $a_{3{\rm ln}}$ are known \cite{Smirnov:2008pn,Smirnov:2009fh,Anzai:2009tm,Smirnov:2010zc,Anzai:2010td,Brambilla:1999qa}.
Specializing to ${\rm SU}(3)$ and $n_f=2$, they are given by (cf.\ \cite{Jansen:2011vv})
\begin{multline}
a_3=27c_1 +\frac{15}{16}c_2+9c_3+\frac{5}{48}c_4-\frac{968981}{729}-8\pi^2\Bigl(15-\frac{8\pi^2}{45}\Bigr) \\
+144\pi^2\Bigl(\ln3+\gamma_E\Bigr) +\frac{38192}{27}\zeta(3)+\frac{320}{9}\zeta(5)
\end{multline}
and $a_{3{\rm ln}}=144\pi^2$.
The constants $c_i$ ($i=1\ldots4$) are only known numerically:
\begin{equation}
 c_1 = 502.24(1)\,, \quad
 c_2 = -136.39(12)\,, \quad
 c_3 = -709.717\,, \quad
 c_4 = -56.83(1)\,.
\end{equation}
$c_1$ and $c_2$ have been determined independently by both \cite{Smirnov:2009fh} and \cite{Anzai:2009tm}.
We use the numerical values from \cite{Smirnov:2009fh}, who provide smaller statistical errors.\footnote{The errors associated with $c_i$ $(i=1\ldots4)$ turn out to be negligible in the context of our $\Lambda_{\overline{\textrm{\rm MS}}}$ determination; therefore, we will not discuss them any further.} $c_3$ and $c_4$ have been determined by \cite{Smirnov:2008pn}.
The coefficients $a_i$ ($i=1\ldots4$) for arbitrary values of $n_f$ can be found, e.g., in \cite{Jansen:2011vv}.

Therewith, all coefficients in the perturbative expansion of the static potential in momentum space up to order $\alpha_s^4$ have been assembled.
We emphasize again that the resulting expression is independent of the explicit choice for $\mu$, in the sense that 
different choices for $\mu$ only lead to deviations at ${\cal O}(\alpha_s^5)$ [cf.\ \Eqref{eq:RG_alphaV}], which is beyond the accuracy of the contributions taken into account in \Eqref{eq:pert1}.

In order to prevent the logarithms $L$ in \Eqref{eq:pert1} from becoming large and thereby spoil the perturbative expansion, it is desirable to ensure that $\mu$ does not deviate much from $p$.
Hence, a particularly convenient choice for the renormalization scale is $\mu\equiv p$, implying $L=0$. We will exclusively stick to this choice throughout the remainder of this paper.
Adopting this choice, $\alpha_V$ simplifies significantly. It becomes a function of $\alpha_s(p)$ only and reads
\begin{equation}
 \alpha_{V}[\alpha_s(p)] = \alpha_s(p)\left\{1+\frac{\alpha_s(p)}{4\pi}\,a_1
+\left(\frac{\alpha_s(p)}{4\pi}\right)^2a_2
+\left(\frac{\alpha_s(p)}{4\pi}
\right)^3\Bigl[a_3+a_{3 {\rm
ln}}\ln{\alpha_s}(p)\Bigr]+\ldots
\right\} . \label{eq:pert2}
\end{equation}
For completeness, note that -- as briefly mentioned in the introduction -- an analogous choice in position space, i.e.\ setting $\mu\equiv1/r$ completely spoils the convergence of the perturbative potential $V(r)$ \cite{Billoire:1979ih}.
As will also become obvious below, the choice $\mu\equiv p$ does not lead to any problems, the reason being the much less pronounced IR sensitivity of the perturbative potential $V(p)$ in momentum space \cite{Beneke:1998rk}.

Let us here also mention the papers by \cite{Brodsky:1982gc,Lepage:1992xa} who argue that a particularly convenient choice of the RG scale $\mu$ is given by $\mu=p\,{\rm e}^{-5/6}$, rendering \Eqref{eq:pert1} independent of $n_f$ up to ${\cal O}(\alpha_s^2)$. While this choice is especially convenient when comparing the results for $\alpha_V$ as an expansion in powers of $\alpha_s(\mu)$ for different numbers of dynamical quarks as the $n_f$ dependency is relegated to higher order expansion coefficients, for our analysis this choice has no advantages and does not provide a handle to improve the results:
Even though the expansion coefficients of powers of $\alpha_s^{1+n}(\mu)$ show a slightly less pronounced increase with $n$ for this choice (cf.\ also our detailed analysis for $\mu=p$ presented below),
for a given value of $p$ the explicit numerical value of $\alpha_s(p\,{\rm e}^{-5/6})$ is substantially increased as compared to $\alpha_s(p)$.
In the momentum regime where both lattice simulations and perturbative calculations for the $Q\bar Q$ static potential are viable, the combined effect of these two opposite tendencies does not favor $\mu=p\,{\rm e}^{-5/6}$ in comparison to $\mu=p$.

An alternative representation of \Eqref{eq:pert2} is
\begin{multline}
 \frac{\alpha_{V}[\alpha_s(p)]}{\alpha_s(p)} = 1+x\,a_1
+\left(x\,a_1\right)^2\frac{a_2}{a_1^2} \\
+\left(x\,a_1
\right)^3\frac{a_3+a_{3 {\rm
ln}}\ln(4\pi/a_1)}{a_1^3}\Bigl[1+\frac{a_{3 {\rm
ln}}}{a_3+a_{3 {\rm
ln}}\ln(4\pi/a_1)}\ln(xa_1)\Bigr]+\ldots , \label{eq:pert3}
\end{multline}
where we employed the shortcut notation $x\equiv\frac{\alpha_s(p)}{4\pi}$.

In order to allow for more insights into the structure of $\alpha_V[\alpha_s(p)]$ for $n_f=2$, we insert the explicit numerical values for the coefficients $a_1$, $a_2$, $a_3$ and $a_{3{\rm ln}}$ into \Eqref{eq:pert3}, and rewrite it as
\begin{equation}
 \frac{\alpha_{V}[\alpha_s(p)]}{\alpha_s(p)} \approx 1+x\,a_1
+5.00\left(x\,a_1\right)^2
+17.63\left(x\,a_1
\right)^3\Bigl[1+0.15\ln(xa_1)\Bigr]+\ldots , \label{eq:pert4}
\end{equation}
with $a_1\approx8.11$.
It is easy to check that for $xa_1\lesssim\frac{1}{5}$ $\leftrightarrow$ $\alpha_s(p)\lesssim\frac{4\pi}{5a_1}\approx0.31$ the contributions $\sim(xa_1)^n$ in \Eqref{eq:pert4} are ordered in the sense that they become increasingly less important, when increasing $n$ from $0$ to $3$.
For larger values of $\alpha_s(p)\gtrsim0.31$ this ordering is spoiled.
Correspondingly, the perturbative expansion of $\alpha_V[\alpha_s(p)]$ with $n_f=2$ can in particular be considered as well-behaved and controllable for $\alpha_s(p)\ll 0.31$.
As the highest order contribution  $\sim(xa_1)^3$ in \Eqref{eq:pert4} is still significantly smaller than the next-to-highest one $\sim(xa_1)^2$, in the present paper we will consider the perturbative expansion as trustworthy even up to $\alpha_s(p)\lesssim 0.31$.
Taking into account the value of $\Lambda_{\overline{\rm MS}}$ as determined in \cite{Jansen:2011vv}, $\Lambda_{\overline{\rm MS}}^{(n_f=2)}=315 \, \textrm{MeV}$, this directly translates into a restriction to momenta $p \gtrsim 1500 \, \textrm{MeV}$ (cf.\ Sec.~\ref{sec:LambdaMSbar}).


\subsection{\label{sec:LambdaMSbar}The scale $\Lambda_{\overline{\rm MS}}$}

So far we have not discussed, how an explicit numerical value can be attributed to the strong coupling $\alpha_s(\mu)$ evaluated at a given momentum $\mu$.
We emphasize that such an identification is renormalization scheme dependent, the most widely used scheme being the $\overline{\rm MS}$ scheme \cite{Bardeen:1978yd,Furmanski:1981cw}, which we also adopt here.

A straightforward integration of \Eqref{eq:beta} yields
\begin{equation}
 \ln\frac{\mu}{\Lambda}
 =\biggl[\int\frac{{\rm d}\alpha_s}{\alpha_s}\frac{1}{\beta(\alpha_s)}\biggr]\bigg|_{\alpha_s=\alpha_s(\mu)} + C\,, \label{eq:beta2}
\end{equation}
where the conventional definition of $\Lambda$ in the $\overline{\rm MS}$ scheme, i.e.\ $\Lambda\to\Lambda_{\overline{\rm MS}}$,
corresponds to the choice of $C=\frac{\beta_1}{2\beta_0^2}\ln(\frac{\beta_0}{4\pi})$ \cite{Chetyrkin:1997sg}\footnote{For completeness, note that our conventions slightly differ from those of \cite{Chetyrkin:1997sg},
who write \Eqref{eq:beta} in terms of $a\equiv\alpha_s/\pi$. Moreover, $\beta_n|_{\textrm{\cite{Chetyrkin:1997sg}}}=\beta_n/4^{n+1}$ and $b_n|_{\textrm{\cite{Chetyrkin:1997sg}}}=\beta_n/(4^n\beta_0)$.}.

Employing some elementary manipulations and rearrangements, \Eqref{eq:beta2} adopted to the $\overline{\rm MS}$ scheme can be written as
\begin{equation}
 \Lambda_{\overline{\rm MS}}\equiv\mu\left(\frac{\beta_0\alpha_s(\mu)}{4\pi}\right)^{-\frac{\beta_1}{2\beta_0^2}}\exp\left\{-\frac{2\pi}{\beta_0\alpha_s(\mu)}-\int_0^{\alpha_s(\mu)}\frac{{\rm d}\alpha_s}{\alpha_s}\left[\frac{1}{\beta(\alpha_s)}+\frac{2\pi}{\beta_0\alpha_s}-\frac{\beta_1}{2\beta_0^2}\right]\right\}, \label{eq:Lambda2-4}
\end{equation}
involving a definite integral over $\alpha_s$ (cf.\ e.g.\ \cite{Capitani:1998mq}). The additional terms apart from the factor of $1/\beta(\alpha_s)$ have been included in the integrand to make the finiteness of the integral over the interval from $0$ to $\alpha_s(\mu)$ manifest.  

In general \Eqref{eq:Lambda2-4} cannot be solved explicitly to provide the coupling $\alpha_s(\mu)$ at a given momentum scale $\mu$ as a function of the ratio $\mu/\Lambda_{\overline{\rm MS}}$. An exact closed form solution is only possible at leading order -- i.e.\ taking into account only the leading contribution of the $\beta$-function~\eqref{eq:betaseries}, $\beta(\alpha_s)\approx-\frac{\alpha_s}{2\pi}\beta_0$ -- and reads $\alpha_s(\mu)=\bigl[\frac{\beta_0}{2\pi}\ln(\mu/\Lambda)\bigr]^{-1}$.
Note, however, that approximate results for $\alpha_s(\mu)$ as a function of the ratio $\mu/\Lambda_{\overline{\rm MS}}$ are available also for higher order contributions:
the derivation of such expressions involves expansions in terms of $\ln(\mu/\Lambda_{\overline{\rm MS}})$; see e.g.\ \cite{Chetyrkin:1997sg,Beringer:1900zz}.

Aiming at the determination of $\Lambda_{\overline{\rm MS}}$ by fitting the perturbative expression for the static potential~\eqref{eq:V(p)} to numerical data from lattice simulations
we do not see any reason to resort to these further approximations. Our strategy is rather to solve the implicit equation~\eqref{eq:Lambda2-4} for $\alpha_s(\mu)$ numerically. 
Hence, we will always use the relation~\eqref{eq:Lambda2-4} between $\alpha_s(\mu)$ and $\mu/\Lambda_{\overline{\rm MS}}$ at the presently best available accuracy, irrespectively of the order of the expansion of the perturbative potential in $\alpha_s(p)$.

However, note that there is still some freedom left in deciding, how to proceed with the evaluation of \Eqref{eq:Lambda2-4}.
We can
\begin{itemize}
 \item[(I)] either plug in the perturbative expression of the $\beta$-function~\eqref{eq:betaseries} at the presently best known accuracy and then do the integration over $\alpha_s$ numerically,
 \item[(II)] or adopt a Taylor expansion of the integrand in \Eqref{eq:Lambda2-4} and do the integral analytically, keeping only those terms, whose coefficients are known explicitly,
\begin{multline}
 \int_0^{\alpha_s(\mu)}\frac{{\rm d}\alpha_s}{\alpha_s}\left[\frac{2}{\beta(\alpha_s)}+\frac{4\pi}{\beta_0\alpha_s}-\frac{\beta_1}{\beta_0^2}\right] \\
 = \frac{\beta_0\beta_2 - \beta_1^2}{\beta_0^3}\frac{\alpha_s(\mu)}{4\pi} + \frac{\beta_0^2\beta_3 - 2\beta_0\beta_1\beta_2+\beta_1^3}{2\beta_0^4}\left(\frac{\alpha_s(\mu)}{4\pi}\right)^2 +{\cal O}(\alpha_s^3). \label{eq:auch}
\end{multline}
\end{itemize}
Of course, limiting ourselves to the truly perturbative regime, i.e.\ the regime, where the perturbative expansion of the $\beta$-function~\eqref{eq:betaseries} is such that higher order corrections become increasingly less important,
both choices should be equally justified. 

As pointed out in Sec.~\ref{sec:perturbation}, when fitting the perturbative static potential~\eqref{eq:V(p)} in momentum space to lattice data we will always stick to the identification $\mu\equiv p$.
Correspondingly, higher order corrections in \Eqref{eq:betaseries} should remain small throughout the integration interval from $0$ to $\alpha(p)$ in \Eqref{eq:Lambda2-4} and \Eqref{eq:auch}, respectively. 

To get a feeling, how the perturbative expansion of the $\beta$-function behaves,
we employ the same strategy as adopted in the context of \Eqref{eq:pert4} and study its behavior at the upper integration limit in \Eqref{eq:Lambda2-4}, i.e.\ at $\alpha_s\equiv\alpha_s(p)$.
First we rewrite \Eqref{eq:betaseries} as
\begin{equation}
 \beta(\alpha_s)=-\frac{\alpha_s}{2\pi}\beta_0\biggl[1+x\,\frac{\beta_1}{\beta_0}+\biggl(x\,\frac{\beta_1}{\beta_0}\biggr)^2\frac{\beta_0\beta_2}{\beta_1^2}+\biggl(x\,\frac{\beta_1}{\beta_0}\biggr)^3\frac{\beta_0^2\beta_3}{\beta_1^3}
+ \ldots\,\biggr] \label{eq:betaseries2}
\end{equation}
with $x\equiv\frac{\alpha_s(p)}{4\pi}$. Second we insert the explicit numerical coefficients of the $\beta$-function for $n_f=2$ into this equation, resulting in
\begin{equation}
 \beta(\alpha_s)\approx-\frac{\alpha_s}{2\pi}\beta_0\biggl[1+x\,\frac{\beta_1}{\beta_0}+1.47\biggl(x\,\frac{\beta_1}{\beta_0}\biggr)^2+3.52\biggl(x\,\frac{\beta_1}{\beta_0}\biggr)^3
+ \ldots\,\biggr] \label{eq:betaseries3}
\end{equation}
with $\frac{\beta_1}{\beta_0}\approx7.93$.
Using the same reasoning as in the context of \Eqref{eq:pert4} above, we find that for the contributions $\sim(x\beta_1/\beta_0)^n$ to become increasingly less important with $n$,
we have to demand $x\beta_1/\beta_0\lesssim0.4$, which corresponds to $\alpha_s(p)\lesssim0.63$.

Hence, in particular for $\alpha_s(p)\lesssim0.31$ -- which was the value found to crudely delimit the range of validity of a perturbative expansion of the static potential in momentum space for $n_f=2$ [cf.\ in the context of \Eqref{eq:pert4} above] -- higher order terms in the perturbative expansion of the $\beta$-function~\eqref{eq:betaseries} for $n_f=2$ are expected to become much less important.
In turn both possibilities (I) and (II) discussed above to numerically solve \Eqref{eq:Lambda2-4} for $\alpha_s(p)$ as a function of $p/\Lambda_{\overline{\rm MS}}$ should yield very similar results.


\newpage

\section{\label{SEC600}Determination of $\Lambda_{\overline{\rm MS}}$}

In this section, we determine $\Lambda_{\overline{\textrm{MS}}}$ in units of the lattice spacing, i.e.\ $a\Lambda_{\overline{\textrm{MS}}}$, by fitting perturbative expressions for the static potential in momentum space (cf.\ Sec.~\ref{sec:perturbation}) to corresponding lattice results (cf.\ Sec.~\ref{SEC572}). Using the values of the lattice spacing listed in Table~\ref{TAB077}, these results can easily be converted to physical units, i.e.\ $\textrm{MeV}$. Unless explicitly stated otherwise, the errors provided for $\Lambda_{\overline{\textrm{MS}}}$ do not account for the uncertainties associated with the lattice spacing $a$. These errors will however be accounted for when quoting our final result for $\Lambda_{\overline{\textrm{MS}}}$ [cf.\ \Eqref{EQN844}, below].
For completeness, we will also quote $r_0\Lambda_{\overline{\textrm{MS}}}$, which is dimensionless and, hence, unaffected by any potential uncertainty in $a$ (uncertainties in $r_0 / a$, which are collected in Table~\ref{TAB077}, are, of course, included).


\subsection{Fitting procedures}

The perturbative $Q\bar Q$ potential to be fitted to lattice results is given by
\begin{equation}
\label{EQN678} V(p) = -\frac{4}{3}\frac{4\pi}{p^2}\alpha_s(p) \biggl\{
\underbrace{\underbrace{\underbrace{\underbrace{1}_{\textrm{LO}}
+ \frac{\alpha_s(p)}{4 \pi} a_1}_{\textrm{NLO}}
+ \bigg(\frac{\alpha_s(p)}{4 \pi}\bigg)^2 a_2}_{\textrm{NNLO}}
+ \bigg(\frac{\alpha_s(p)}{4 \pi}\bigg)^3 \Bigl[a_3 + a_{3 {\rm ln}} \ln{\alpha_s(p)}\Bigr]}_{\textrm{NNNLO}}
\biggr\}+V_0,
\end{equation}
where we included an overall constant offset $V_0$ of the potential.
We use different orders in the expansion of $V(p)$ in powers of $\alpha_s(p)$ to test and judge the convergence behavior of our results; the abbreviation ${\rm N}^n{\rm LO}$ stands for (next-to-)$^n$leading-order. 
As only energy differences are measurable $V_0$ does not have any observable consequences. However, it is necessary to allow for a meaningful matching of the perturbative potential to lattice results. 

For the relation between $\alpha_s(p)$ and $\Lambda_{\overline{\rm MS}}$ we employ (cf.\ Sec.~\ref{sec:perturbation})
\begin{itemize}
\item[(I)] either \Eqref{eq:Lambda2-4} with $\beta(\alpha)$ given by the terms written explicitly in \Eqref{eq:betaseries},
\item[(II)] or \Eqref{eq:Lambda2-4} with the integral expression substituted for the terms written explicitly on the left-hand side of \Eqref{eq:auch}.
\end{itemize}
These implicit equations are solved numerically to yield $\alpha_s(p)$ as a function of $p/\Lambda_{\overline{\rm MS}}$.

We employ an uncorrelated $\chi^2$ minimizing fit with two degrees of freedom, $V_0$ and $\Lambda_{\overline{\rm MS}}$, to fit the perturbative $Q\bar Q$ potential~\eqref{EQN678} to the lattice potential in momentum space.
Our fitting interval is delimited by $p_{\textrm{min}}$ and $p_{\textrm{max}}$.
Note that the lattice potential in momentum space originates from position space results whose large distance behavior is modeled by \Eqref{EQN600} with $M \in \{1,2,3,4\}$ -- and thus also depends on $M$ (cf. Sec.~\ref{SEC572}).


\subsection{\label{SEC329}Systematic and statistical errors of $\Lambda_{\overline{\rm MS}}$}


\subsubsection{\label{input_var}Individual variation of input parameters}

We investigate the stability of $\Lambda_{\overline{\rm MS}}$ with respect to variations of the input parameters $p_\textrm{min}$, $p_\textrm{max}$ and $M$, using our finest lattice spacing $a \approx 0.042 \, \textrm{fm}$ and $(L'/a)^3 = 256^3$ lattice sites. To this end the input parameters delimiting the fitting range are varied in the following intervals:
\begin{itemize}
\item $p_\textrm{min} = 1500 \, \textrm{MeV} \ldots 2250 \, \textrm{MeV}$: \\
As discussed in detail in Sec.~\ref{sec:perturbation}, for $p_\textrm{min} \lesssim 1500 \, \textrm{MeV}$ the validity of perturbative expressions for the static potential is rather questionable.

\item $p_\textrm{max} = 2250 \, \textrm{MeV} \ldots 3000 \, \textrm{MeV}$: \\
The maximum momentum on our finest lattice along an axis is $\overline{p} = \pi / a \approx 15 \, \textrm{GeV}$. For $p\lesssim \overline{p} / 3$, it seems reasonable to expect rather small discretization errors.
These expectations have also been confirmed by numerical investigations.
Fitting the perturbative potential to lattice results we obtain an essentially stable value of $\Lambda_{\overline{\rm MS}}$ up to $3000 \, \textrm{MeV} \ldots 3500 \, \textrm{MeV}$;
cf.\ also Figure~\ref{FIG003} (c) and (d) below.
\end{itemize}
The different choices for the parameter $M$ are
\begin{itemize}
\item $M\in\{1,2,3,4\}$: \\
The parameter $M$ is varied to estimate the systematic errors associated with our modeling of the long range part of the lattice potential in position space; cf.\ Sec.~\ref{EQN808}.
\end{itemize}
Below, we demonstrate that the fit results for $\Lambda_{\overline{\rm MS}}$ are rather stable with respect to these parameter variations, i.e.\ we confirm that a meaningful and rather precise matching of perturbative expressions for the $Q\bar Q$ static potential and lattice results is possible.

\begin{figure}[htb]
  \centering
   \subfigure[LO - $\Lambda_{\overline{\rm MS}}=534(3)$MeV]{\includegraphics[width=0.4\textwidth]{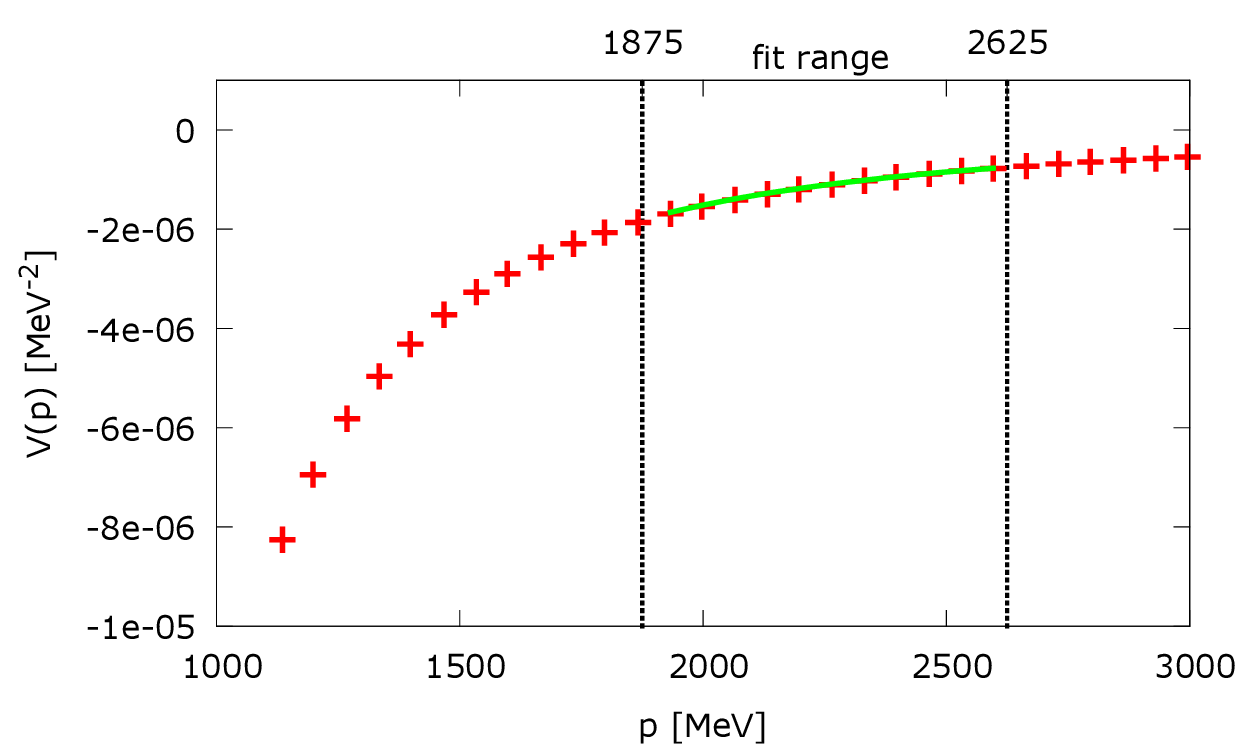}}\qquad
   \subfigure[NLO - $\Lambda_{\overline{\rm MS}}=414(3)$MeV ]{\includegraphics[width=0.4\textwidth]{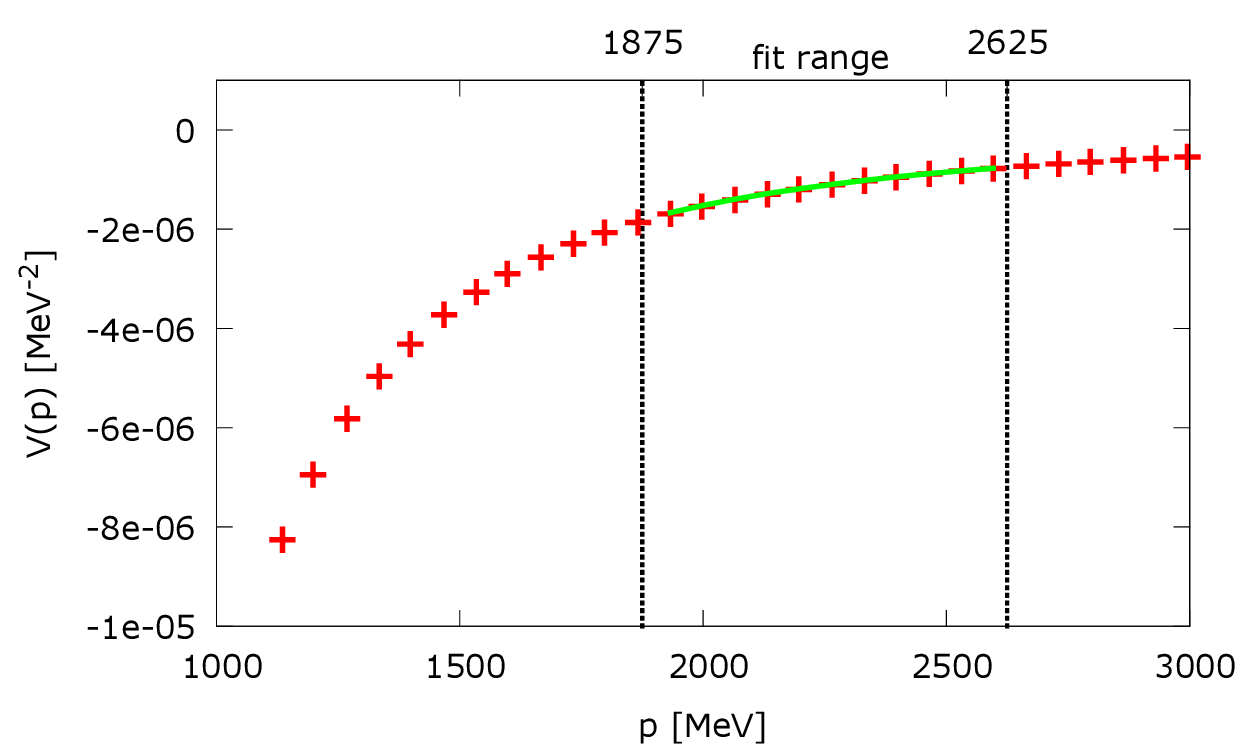}}\\
   \subfigure[NNLO - $\Lambda_{\overline{\rm MS}}=339(2)$MeV ]{\includegraphics[width=0.4\textwidth]{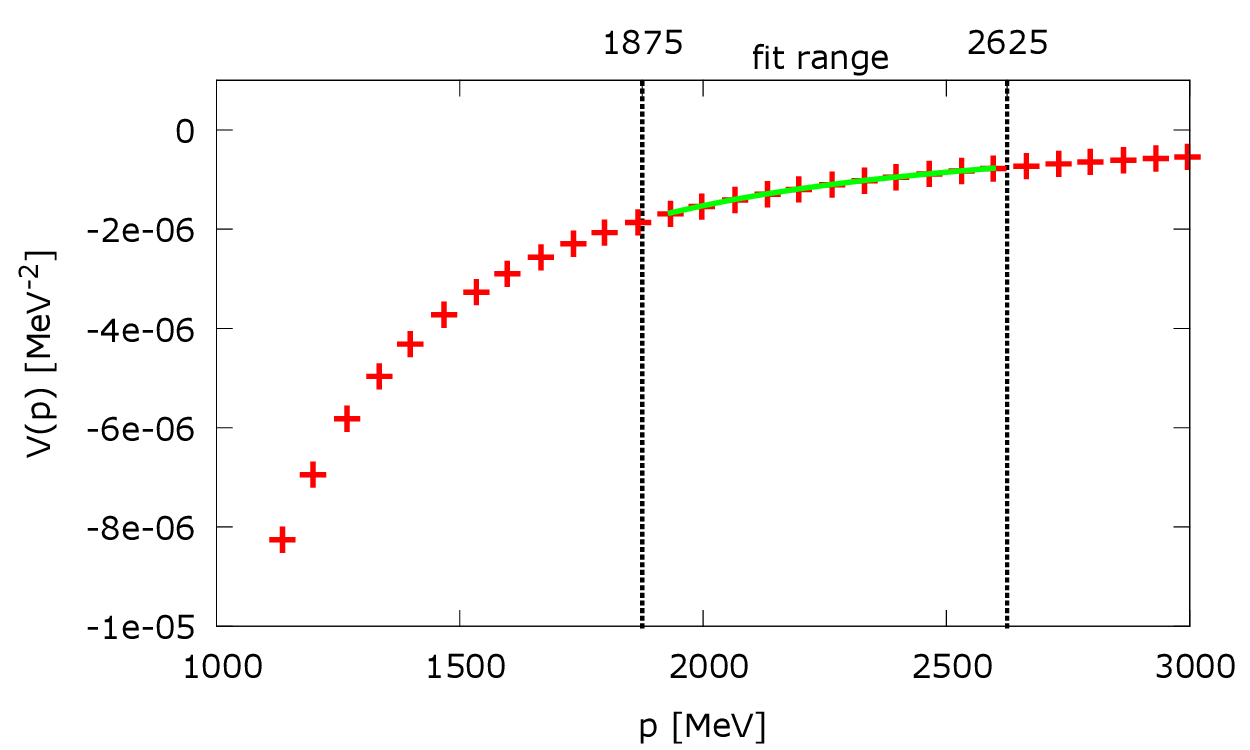}}\qquad
   \subfigure[NNNLO - $\Lambda_{\overline{\rm MS}}=315(2)$MeV ]{\includegraphics[width=0.4\textwidth]{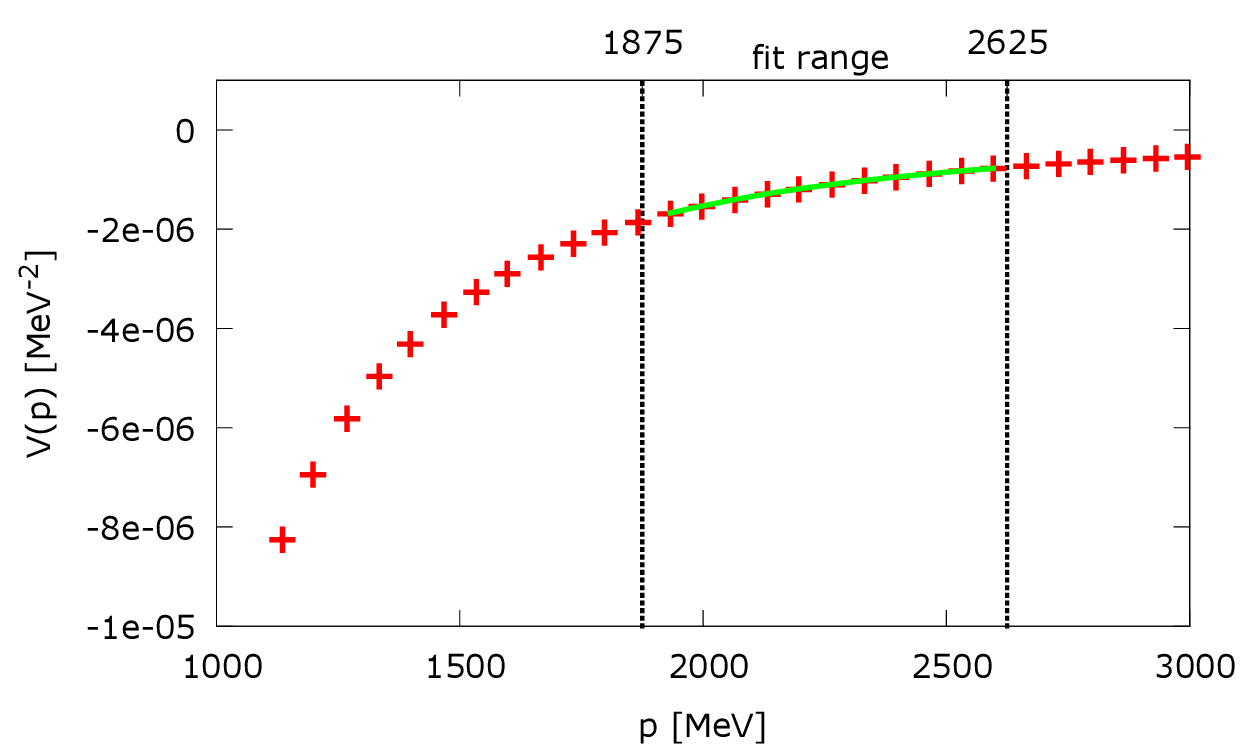}}
  \caption{\label{FIG002}Exemplary fits of the perturbative static potential~\eqref{EQN678} to lattice results obtained with $\beta = 4.35$ and $M = 3$. We employ fitting procedure (I) with $p_\textrm{min} = 1875 \, \textrm{MeV}$ and $p_\textrm{max} = 2625 \, \textrm{MeV}$. The errors quoted below the plots refer to the fitting and are statistical errors only; systematic errors will be discussed below.}
\end{figure}

\begin{figure}[p]
\centering
\subfigure[fitting proc.\ (I), $p_{\textrm{max}}=2625 \, \textrm{MeV}$]{\includegraphics[width=0.47\textwidth]{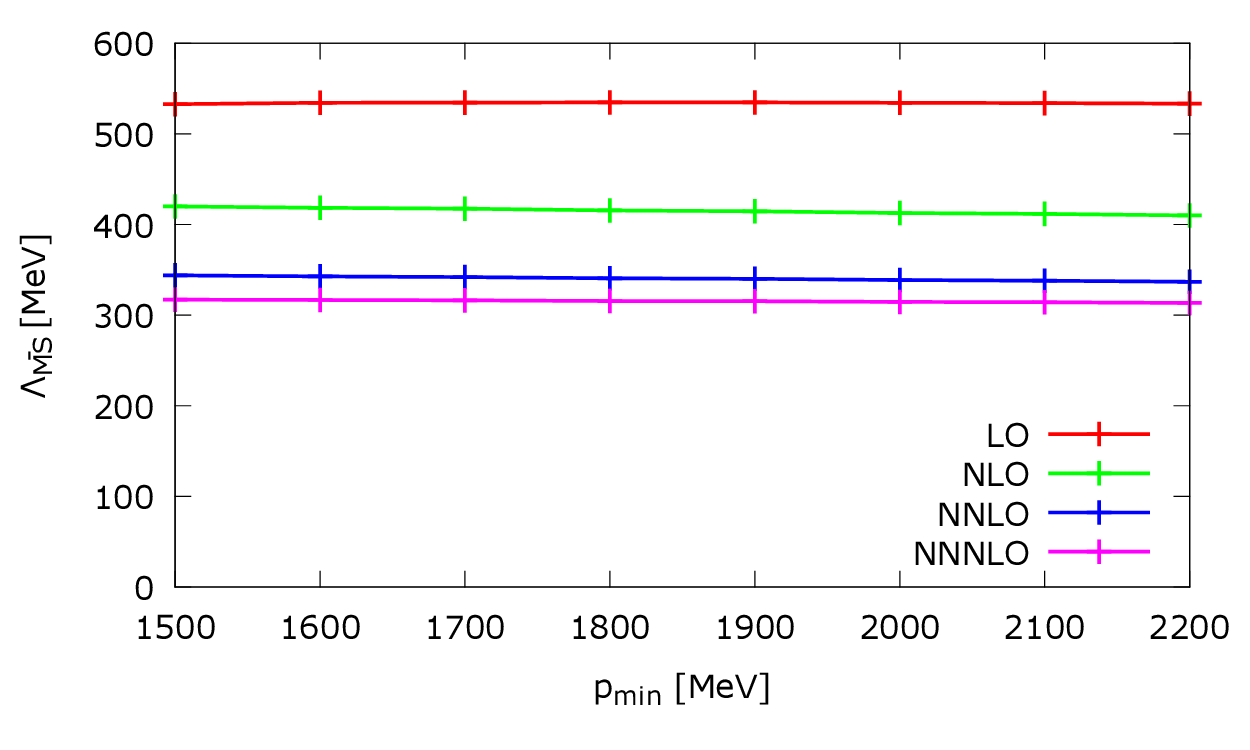}}\qquad
\subfigure[fitting proc.\ (II), $p_{\textrm{max}}=2625 \, \textrm{MeV}$]{\includegraphics[width=0.47\textwidth]{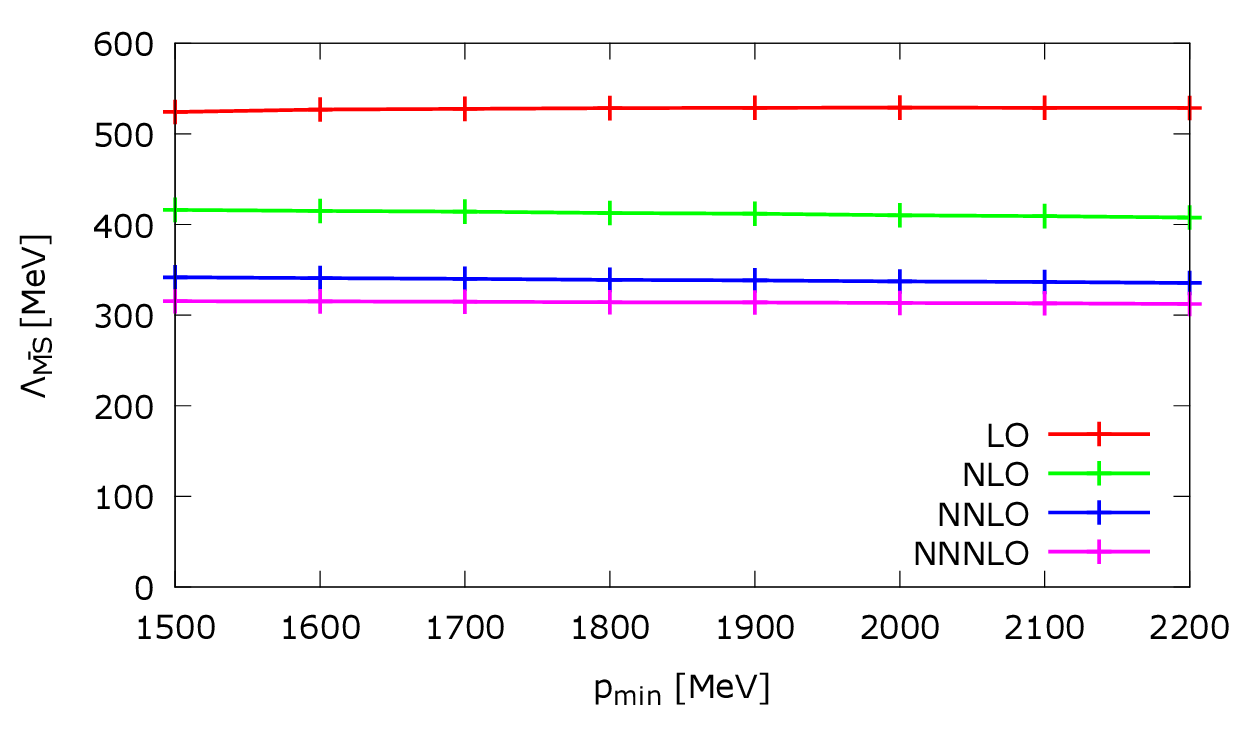}}\\
\subfigure[fitting proc.\ (I), $p_{\textrm{min}}=1875 \, \textrm{MeV}$]{\includegraphics[width=0.47\textwidth]{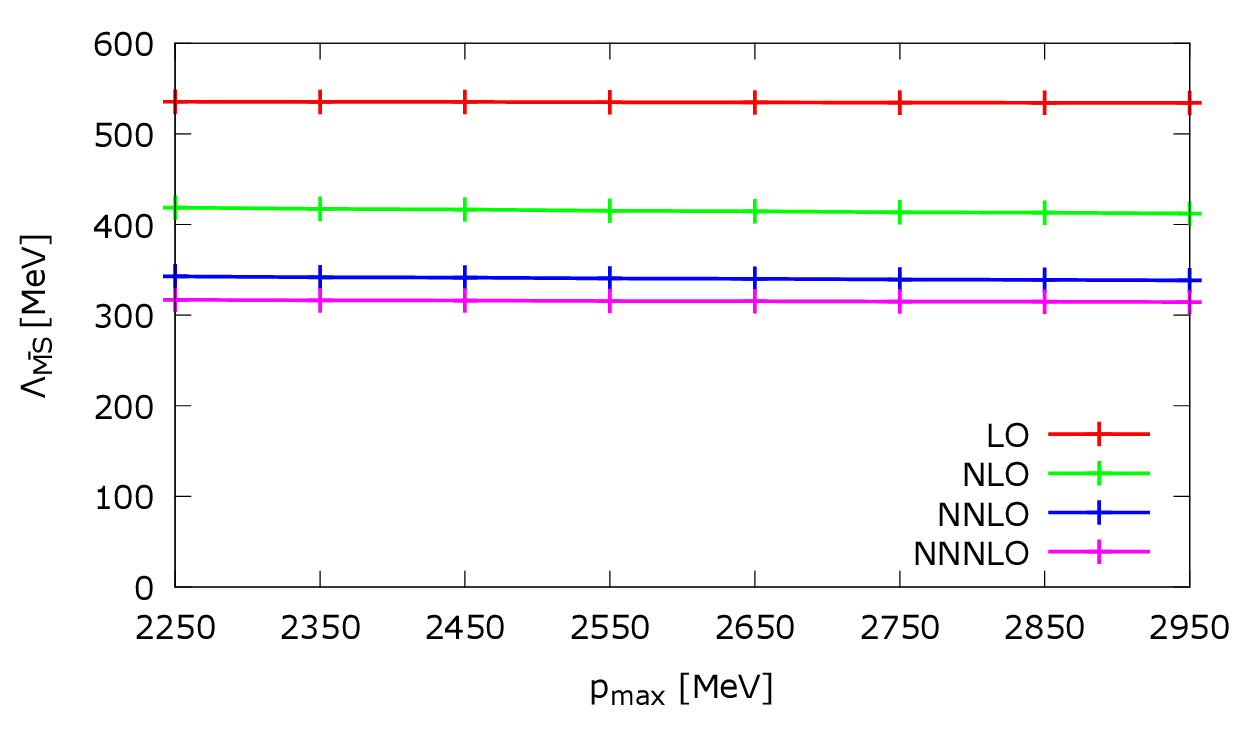}}\qquad
\subfigure[fitting proc.\ (II), $p_{\textrm{min}}=1875 \, \textrm{MeV}$]{\includegraphics[width=0.47\textwidth]{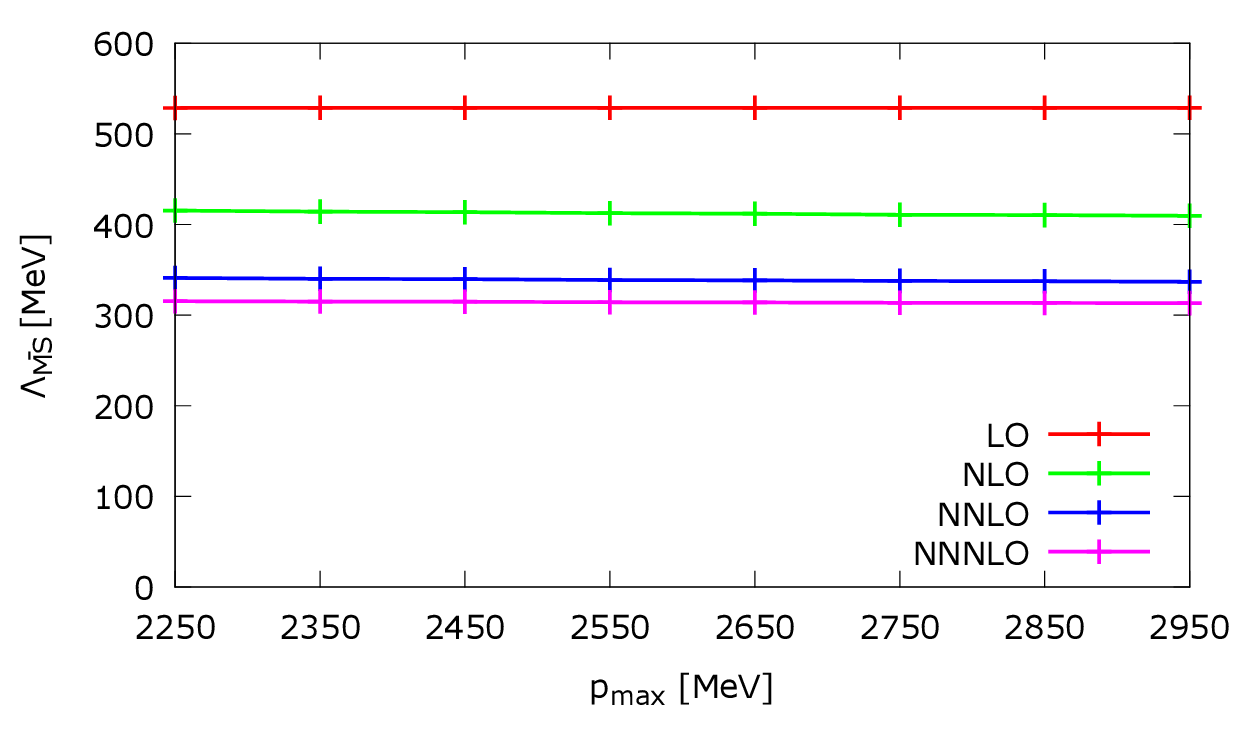}}\\
\subfigure[fitting proc.\ (I), $p_{\textrm{max}}-{p_{\textrm{min}}}=750 \, \textrm{MeV}$]{\includegraphics[width=0.47\textwidth]{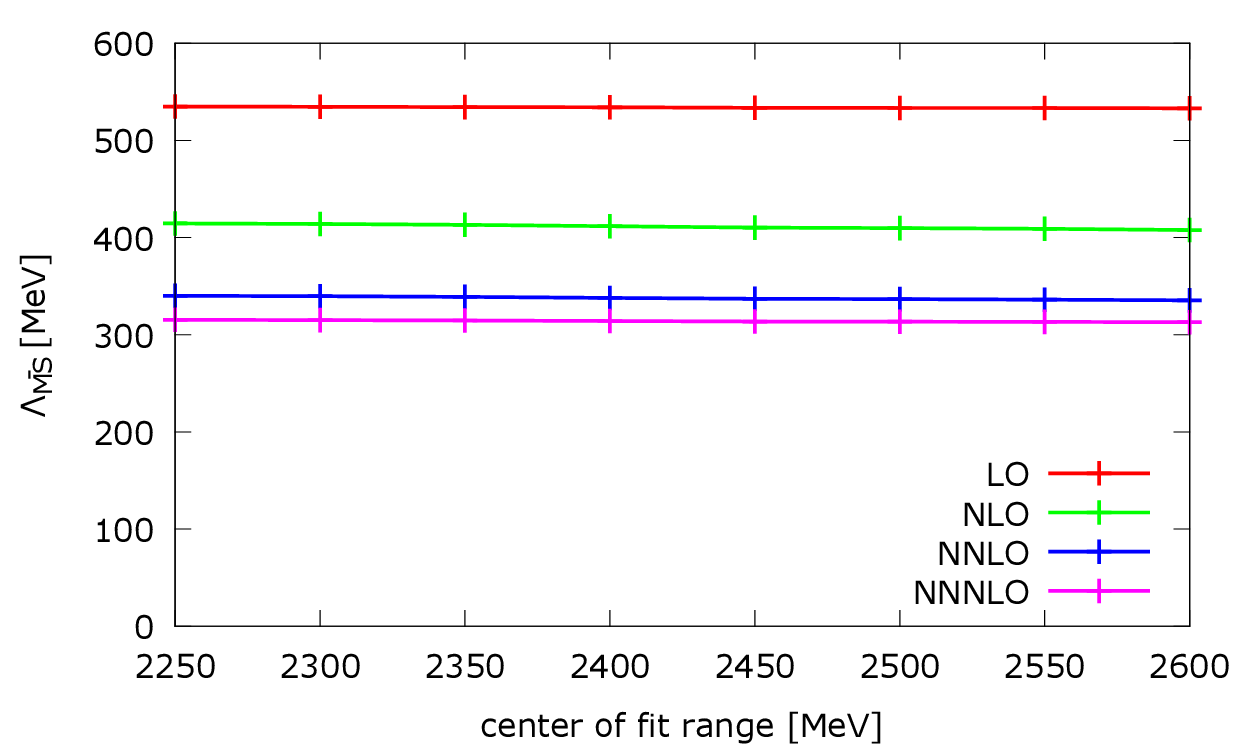}}\qquad
\subfigure[fitting proc.\ (II), $p_{\textrm{max}}-{p_{\textrm{min}}}=750 \, \textrm{MeV}$]{\includegraphics[width=0.47\textwidth]{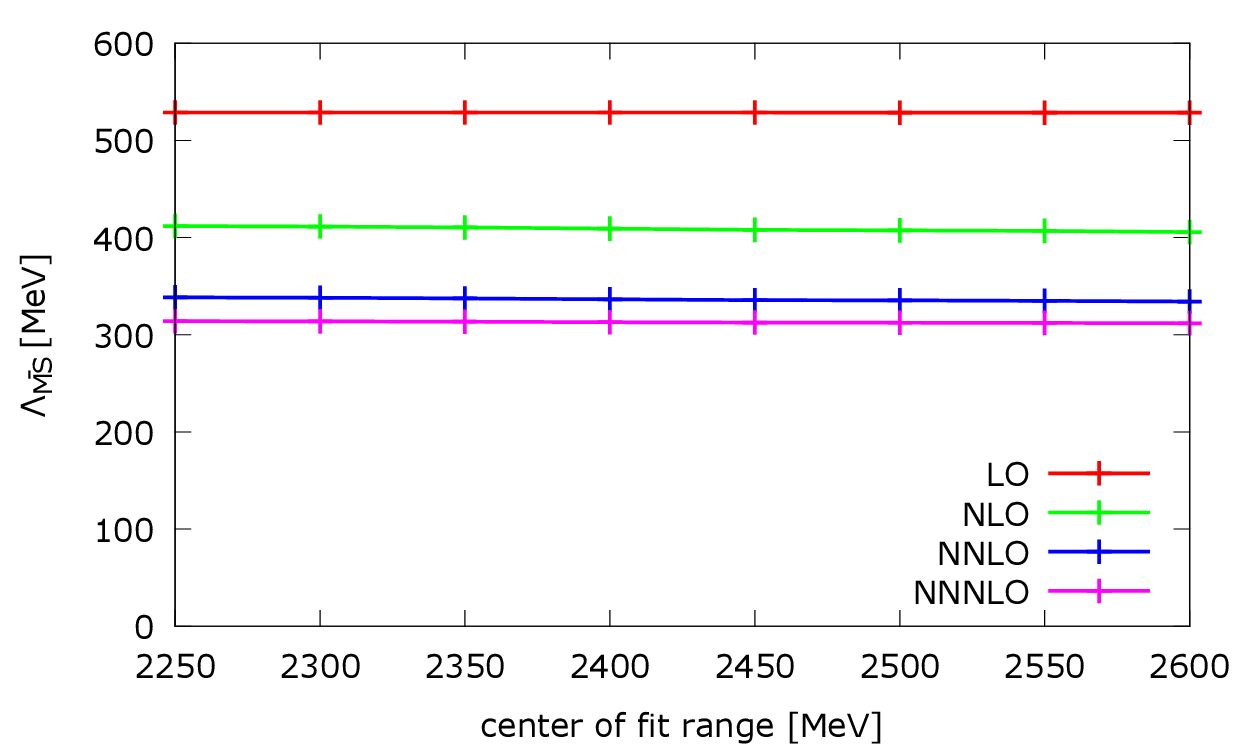}}
\caption{\label{FIG003}Results for $\Lambda_{\overline{\rm MS}}$ obtained by fitting \Eqref{EQN678} to lattice results using $M = 3$; \textbf{left column:} fitting procedure (I); \textbf{right column:} fitting procedure (II); \textbf{first line:} $\Lambda_{\overline{\rm MS}}$ as a function of $p_\textrm{max}$; \textbf{second line:} $\Lambda_{\overline{\rm MS}}$ as a function of $p_\textrm{min}$; \textbf{third line:} $\Lambda_{\overline{\rm MS}}$ as a function of $(p_\textrm{min} + p_\textrm{max})/2$.}
\end{figure}

Exemplary fits of the perturbative static potential~\eqref{EQN678} at LO, NLO, NNLO and NNNLO to lattice results are depicted in Figure~\ref{FIG002}.
Noteworthily, all four orders are suited to describe the lattice potential within statistical errors, i.e.\ fulfill $\chi^2 / \textrm{dof} \lesssim 1.0$.
Similar $\chi^2 / \textrm{dof}$ are obtained when varying input parameters (cf. below).

To understand, how the value of $\Lambda_{\overline{\rm MS}}$ depends on the input parameters $p_\textrm{min}$ and $p_\textrm{max}$, we vary them individually.
Our results are summarized graphically in Figure~\ref{FIG003}:\footnote{\label{footnote}Statistical errors associated with the $\chi^2$ minimizing fits, which are rather small ($\approx 2\,\textrm{MeV}\ldots4{\rm MeV}$), are not considered in this subsection. They are, however, included in the final results presented in Sec.~\ref{SEC790}.}
\begin{itemize}
\item The plots in the left column 
are obtained with fitting procedure (I), while those in the right column 
result from fitting procedure (II).

\item For the plots in the first line 
we vary $p_\textrm{min} = 1500 \, \textrm{MeV} \ldots 2250 \, \textrm{MeV}$, while keeping $p_\textrm{max} = 2625 \, \textrm{MeV}$ fixed at the center of the interval defined above, and set $M=3$. We find:
\begin{itemize}
\item Variation of $\Lambda_{\overline{\rm MS}}$ (NNLO): \\
$337 \, \textrm{MeV} \ldots 344 \, \textrm{MeV}$ (fitting procedure (I)), \\
$335 \, \textrm{MeV} \ldots 342 \, \textrm{MeV}$ (fitting procedure (II)).

\item Variation of $\Lambda_{\overline{\rm MS}}$ (NNNLO): \\
$313 \, \textrm{MeV} \ldots 317 \, \textrm{MeV}$ (fitting procedure (I)), \\
$312 \, \textrm{MeV} \ldots 315 \, \textrm{MeV}$ (fitting procedure (II)).
\end{itemize}

\item For the plots in the second line 
we vary $p_\textrm{max} = 2250 \, \textrm{MeV} \ldots 3000 \, \textrm{MeV}$, while keeping $p_\textrm{min} = 1875 \, \textrm{MeV}$ fixed at the center of the interval defined above, and set $M=3$. We obtain:
\begin{itemize}
\item Variation of $\Lambda_{\overline{\rm MS}}$ (NNLO): \\
$338 \, \textrm{MeV} \ldots 343 \, \textrm{MeV}$ (fitting procedure (I)), \\
$337 \, \textrm{MeV} \ldots 341 \, \textrm{MeV}$ (fitting procedure (II)).

\item Variation of $\Lambda_{\overline{\rm MS}}$ (NNNLO): \\
$314 \, \textrm{MeV} \ldots 317 \, \textrm{MeV}$ (fitting procedure (I)), \\
$313 \, \textrm{MeV} \ldots 315 \, \textrm{MeV}$ (fitting procedure (II)).
\end{itemize}

\item For the plots in the third line 
we vary the center of the fitting range $(p_\textrm{min} + p_\textrm{max}) / 2 = 1875 \, \textrm{MeV} \ldots 2625 \, \textrm{MeV}$, while keeping the width of the fitting range $p_\textrm{max} - p_\textrm{min} = 750 \, \textrm{MeV}$ fixed, and set $M=3$. This results in:
\begin{itemize}
\item Variation of $\Lambda_{\overline{\rm MS}}$ (NNLO): \\
$335 \, \textrm{MeV} \ldots 340 \, \textrm{MeV}$ (fitting procedure (I)), \\
$334 \, \textrm{MeV} \ldots 338 \, \textrm{MeV}$ (fitting procedure (II)).

\item Variation of $\Lambda_{\overline{\rm MS}}$ (NNNLO): \\
$313 \, \textrm{MeV} \ldots 315 \, \textrm{MeV}$ (fitting procedure (I)), \\
$312 \, \textrm{MeV} \ldots 314 \, \textrm{MeV}$ (fitting procedure (II)).
\end{itemize}
\end{itemize}
To allow for a meaningful determination of $\Lambda_{\overline{\rm MS}}$, it is important to ensure that the fit results for $\Lambda_{\overline{\rm MS}}$ only exhibit a weak -- preferably negligible -- dependence on $p_\textrm{min}$ and $p_\textrm{max}$. In plots of $\Lambda_{\overline{\rm MS}}$ as a function of $p_\textrm{min}$ and $p_\textrm{max}$ this should manifest itself in the formation of plateaus. This behavior is clearly visible in Figure~\ref{FIG003} for all six plots and at all orders in the expansion~\eqref{EQN678}.
Moreover, the curves shown in Figure~\ref{FIG003} clearly reveal a convergence behavior when increasing the order in \Eqref{EQN678} from LO to NNNLO: First, $\Lambda_{\overline{\rm MS}}|_{\rm LO}>\Lambda_{\overline{\rm MS}}|_{\rm NLO}>\Lambda_{\overline{\rm MS}}|_{\rm NNLO}>\Lambda_{\overline{\rm MS}}|_{\rm NNNLO}$, second
the ratios of their relative differences scale as
\begin{equation}
 \Lambda_{\overline{\rm MS}}|_{\rm LO}-\Lambda_{\overline{\rm MS}}|_{\rm NLO}\ :\ \Lambda_{\overline{\rm MS}}|_{\rm NLO}-\Lambda_{\overline{\rm MS}}|_{\rm NNLO}\ :\ \Lambda_{\overline{\rm MS}}|_{\rm NNLO} - \Lambda_{\overline{\rm MS}}|_{\rm NNNLO}\ \approx\ 5\ :\ 3\ :\ 1\,.
\end{equation}
Third, the values of $\Lambda_{\overline{\rm MS}}$ extracted from fits of \Eqref{EQN678} at NNLO and NNNLO to lattice results yield quite similar results. This can be considered as indication that the neglected orders beyond NNNLO in \Eqref{EQN678} will not alter the value of $\Lambda_{\overline{\rm MS}}$ significantly.
As an -- rather conservative -- estimate of the systematic error due to the truncation of the perturbative expansion~\eqref{EQN678} at ${\cal O}(\alpha_s^4)$ we take the difference between the NNLO and the NNNLO results for $\Lambda_{\overline{\rm MS}}$ (cf.\ Sec.~\ref{SEC444}).

Another source of uncertainty in our analysis is the modeling~\eqref{EQN600} used to extrapolate the lattice potential computed from Wilson loops to larger $r$. In order to scrutinize this uncertainty, we study the dependence of $\Lambda_{\overline{\rm MS}}$  on $M\in\{ 1, 2, 3, 4\}$, resorting to fitting procedure (I) and keeping $p_\textrm{min} = 1875 \, \textrm{MeV}$ and $p_\textrm{max} = 2625 \, \textrm{MeV}$ fixed. The corresponding results, which are collected in Table~\ref{TAB015}, indicate that $\Lambda_{\overline{\rm MS}}$ is also rather stable with respect to variations of $M$.

\begin{table}[htb]
\begin{center}
\begin{tabular}{|c|c|c|c|c|}
\hline 
 & $V_1(r)$ & $V_2(r)$ & $V_3(r)$ & $V_4(r)$ \\ 
\hline 
NNLO, fitting procedure (I)   & 349(2) & 343(2) & 340(3) & 346(4) \\
NNLO, fitting procedure (II)  & 347(2) & 341(2) & 338(2) & 345(4) \\
\hline 
NNNLO, fitting procedure (I)  & 323(2) & 318(2) & 315(2) & 321(4) \\
NNNLO, fitting procedure (II) & 322(2) & 317(2) & 314(3) & 320(4) \\
\hline
\end{tabular}
\caption{\label{TAB015}Results for $\Lambda_{\overline{\rm MS}}^{(n_f=2)}$ in $\textrm{MeV}$ for different $M \in \{1, 2, 3, 4\}$ using fitting procedure (I) and $p_\textrm{min} = 1875 \, \textrm{MeV}$ and $p_\textrm{max} = 2625 \, \textrm{MeV}$.}
\end{center}
\end{table}

\begin{figure}[htb]
\begin{center}
\includegraphics[width=0.8\textwidth]{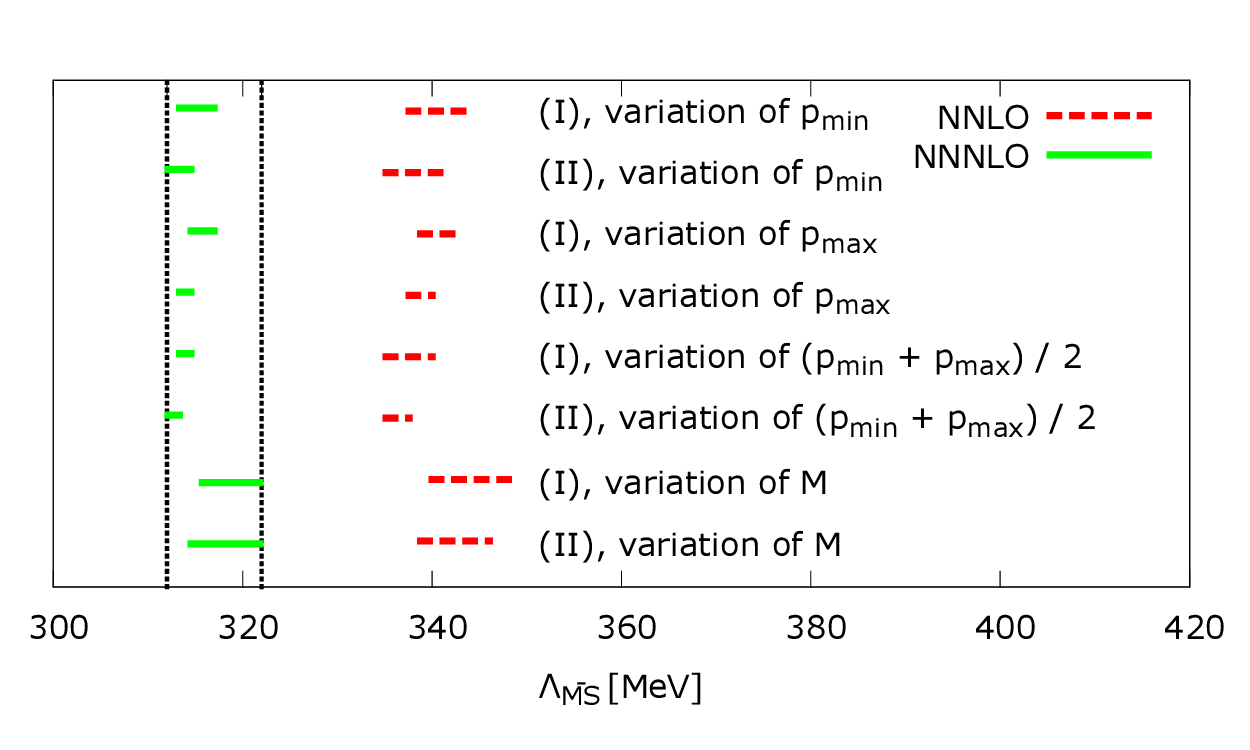}
\end{center}
\caption{\label{FIG003_error}Graphical summary of the variation of $\Lambda_{\overline{\rm MS}}$, when varying the input parameters $p_\textrm{min}$, $p_\textrm{max}$ and $M$ (cf.\ also Figure~\ref{FIG003}, Table~\ref{TAB015} and the corresponding paragraphs in the main text).
For each line, we indicate the fitting procedure used, i.e., either (I) and (II), and specify which parameter is varied. The variation is over the intervals specified at the begin of Sec.~\ref{input_var}.
The other input parameters are held fixed at the center of the respective intervals, and $M=3$ unless otherwise stated.}
\end{figure}

The variation of $\Lambda_{\overline{\rm MS}}$ as a function of the input parameters $p_\textrm{min}$, $p_\textrm{max}$ and $M$ (cf. also Figure~\ref{FIG003} and Table~\ref{TAB015}) when fitting~\eqref{EQN678}, both at NNLO (red) and at NNNLO (green), to lattice results is visualized in Figure~\ref{FIG003_error}. As systematic uncertainty one could e.g.\ quote the whole range of values covered by the NNNLO variations,
\begin{equation}
\label{EQN257} \Lambda_{\overline{\textrm{MS}}}^{(n_f=2)} = 314 \, \textrm{MeV} \ldots 323 \, \textrm{MeV} .
\end{equation}
In particular note the drastic improvement in stability as compared to the previous position space analysis~\cite{Jansen:2011vv}, where a systematic uncertainty larger by a factor of $\approx 4$ has been obtained, $\Lambda_{\overline{\rm MS}}^{(n_f=2)} = 304 \, \textrm{MeV} \ldots 344 \, \textrm{MeV}$ (cf.\ Eq.\ (46) and Figure~5 of \cite{Jansen:2011vv}).

The sources of the systematic error discussed above might, however, be correlated. A method to determine the overall systematic error accounting for potential correlation is discussed in the following subsection.


\subsubsection{\label{SEC444}Consideration of correlations between different systematic error sources}

To account for possible correlations, we perform a large number of fits, with the input parameters chosen randomly and uniformly in the intervals specified in Sec.~\ref{input_var}. As systematic error we then take the variance of the fit results.
This procedure is analogous to that used for a $\Lambda_{\overline{\rm MS}}$ determination from the $Q\bar Q$ static potential in position space based upon the same lattice data \cite{Jansen:2011vv} to which we will compare our results in the following.

We have performed 40,000 fits (i.e., sufficiently many to render the statistical error of the variance negligible):
\begin{itemize}
\item 10,000 NNLO fits, fitting procedure (I);

\item 10,000 NNLO fits, fitting procedure (II);

\item 10,000 NNNLO fits, fitting procedure (I);

\item 10,000 NNNLO fits, fitting procedure (II).
\end{itemize}
In these fits we randomly vary $p_{\rm min}$ and $p_{\rm max}$ in the intervals $p_\textrm{min} = 1500 \, \textrm{MeV} \ldots 2250 \, \textrm{MeV}$ and $p_\textrm{max} = 2250 \, \textrm{MeV} \ldots 3000 \, \textrm{MeV}$ while imposing the constraint $p_\textrm{max} - p_\textrm{min} \geq 375 \, \textrm{MeV}$. Moreover, we cyclically vary $M \in\{ 1, 2, 3, 4\}$. From this analysis we obtain an average and a variance, i.e.\ a systematic error, of
\begin{equation}
\label{EQN791} \Lambda_{\overline{\rm MS}}^{(n_f=2)} =  331(13) \, \textrm{MeV} .
\end{equation}
Again it is most instructive to compare the error with that obtained for a position space analysis of the $Q\bar Q$ static potential based upon the same lattice data \cite{Jansen:2011vv}:
\begin{itemize}
\item The final result of Ref.~\cite{Jansen:2011vv} is $\Lambda_{\overline{\rm MS}}^{(n_f=2)} = 315(26) \, \textrm{MeV}$. Hence, the momentum space analysis pursued in the present paper yields a result which is more precise by a factor of $\approx 2$. 

\item While fitting procedure (A) used in \cite{Jansen:2011vv} is rather similar to (I) and (II) of this paper, fitting procedure (B) used in \cite{Jansen:2011vv} is somewhat different.\footnote{In fact, fitting procedure (B) of \cite{Jansen:2011vv} is expected to increase the error because the one-loop result for $\alpha_s$, employed in this fitting procedure, deviates significantly from the higher-order expressions for $\alpha_s$ throughout the considered fitting interval.} Hence, in order to allow for a consistent and fair comparison of our present momentum space analysis and the previous position space analysis, only fitting procedure (A) should be used. We carried out such an analysis, resulting in $\Lambda_{\overline{\rm MS}}^{(n_f=2)} = 331(20) \, \textrm{MeV}$. While there is now perfect agreement with respect to the average, the error associated with a momentum space analysis is still smaller  by a factor of $\approx 1.5$ than the error obtained in a position space analysis.

\item Finally, one could be less conservative with regard to the estimate of the uncertainty associated with the truncation of the perturbative expressions and only take into account the NNNLO results. A momentum space analysis would then result in $\Lambda_{\overline{\rm MS}}^{(n_f=2)} = 318(3) \, \textrm{MeV}$, while a position space analysis (fitting procedure (A) only) would lead to $\Lambda_{\overline{\rm MS}}^{(n_f=2)} = 326(13) \, \textrm{MeV}$. In this case the error of $\Lambda_{\overline{\rm MS}}$ as obtained from a momentum space analysis is even smaller by a factor of $\approx 4$ as compared to an analogous position space analysis. This rather drastic difference can be attributed to the almost perfect stability of $\Lambda_{\overline{\rm MS}}$ at NNNLO with respect to variations of the fitting range in momentum space (cf.\ Figure~\ref{FIG003}). In position space, a similar stability has not been observed \cite{Jansen:2011vv}.
\end{itemize}
Let us emphasize again that all these results for $\Lambda_{\overline{\rm MS}}^{(n_f=2)}$ agree within statistical errors.

As already mentioned in footnote~\ref{footnote}, the fitting procedure introduces an additional statistical uncertainty of $\approx 2 \, \textrm{MeV} \ldots 4 \, \textrm{MeV}$, which we add in quadrature when generating the final result (cf.\ Sec.~\ref{SEC790}).

Further systematic uncertainties, exclusively originating from the lattice static potential, are discussed in the next section.


\subsubsection{\label{SEC330}Systematic errors of $\Lambda_{\overline{\rm MS}}$ associated with the lattice computation}


\subsubsection*{Lattice discretization errors}

\begin{figure}[htb]
\centering
\includegraphics[width=0.48\textwidth]{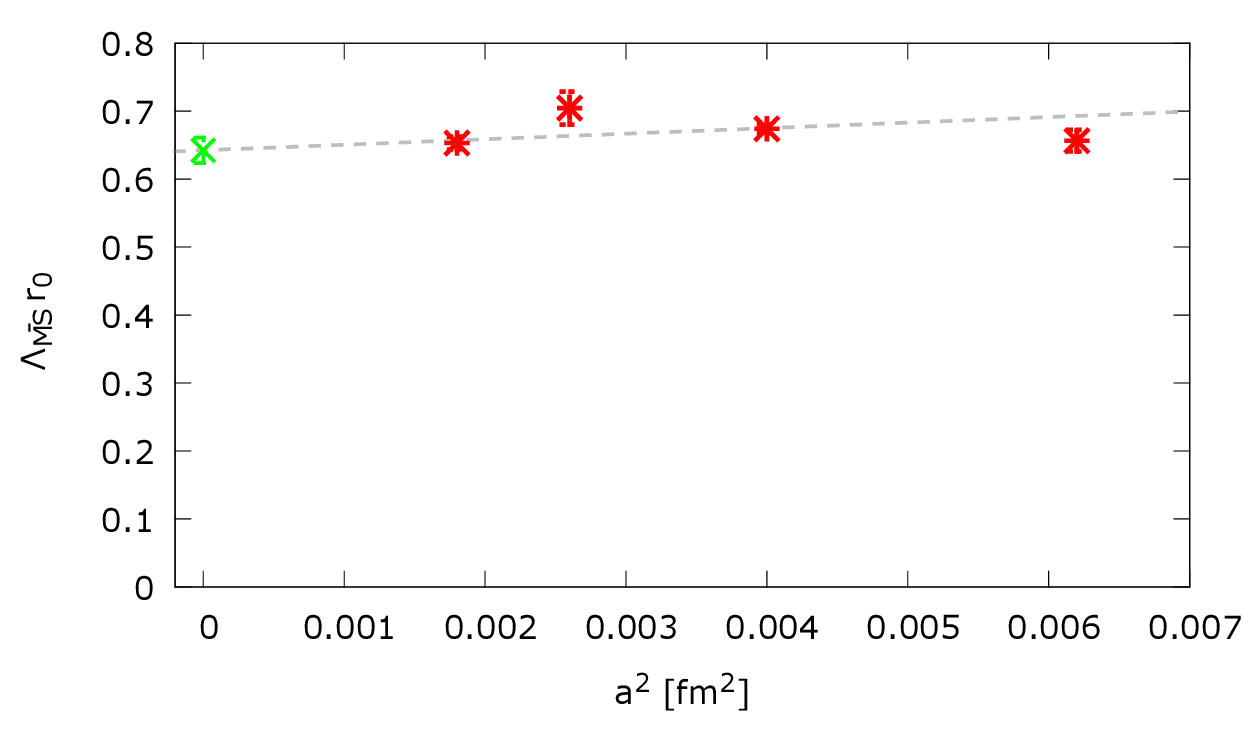}\includegraphics[width=0.48\textwidth]{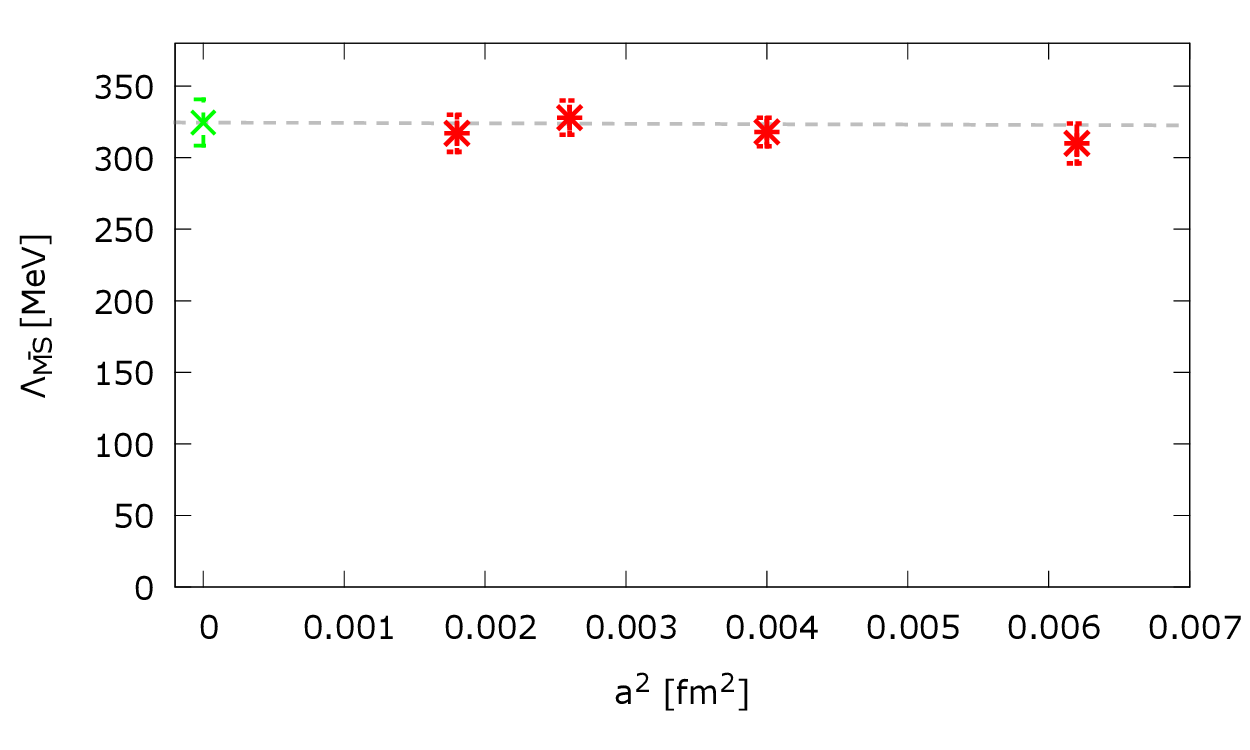} 
\caption{\label{FIG954}Dependence of $r_0\Lambda_{\overline{\rm MS}}$ and $\Lambda_{\overline{\rm MS}}$ on the lattice spacing. The result of an continuum extrapolation is depicted in green.}
\end{figure}

To estimate the order of magnitude of lattice discretization errors, we utilize the lattice ensembles listed in Table~\ref{TAB077}, featuring pions of roughly of the same mass $\approx 284 \, \textrm{MeV} \ldots 352 \, \textrm{MeV}$.
Keeping the input parameters $p_\textrm{min} = 1300 \, \textrm{MeV}$, $p_\textrm{max} = 2050 \, \textrm{MeV}$ and $M=3$ fixed, we extract values for both $r_0\Lambda_{\overline{\rm MS}}$ and  $\Lambda_{\overline{\rm MS}}$ by performing the corresponding fits.
For these investigations we exclusively adopt fitting procedure (I) and use the perturbative $Q\bar Q$ static potential~\eqref{EQN678} at the best known accuracy, i.e.\ NNNLO.\footnote{To allow for stable fits with sufficiently many lattice points for $V(p)$ inside the interval $p_\textrm{min} \ldots p_\textrm{max}$, with $p_\textrm{max}$ not too close to the maximum lattice momentum along one of the three spatial axes, we had to choose $p_\textrm{min}$ slightly smaller than the minimum value of $p_{\rm min}=1500 \, \textrm{MeV}$ defined in Sec.~\ref{input_var}.} In Figure~\ref{FIG954} we depict our results for $r_0\Lambda_{\overline{\rm MS}}$ and $\Lambda_{\overline{\rm MS}}$ as a function of $a^2$. The results determined at the four available lattice spacings perfectly agree within errors, i.e.\ there is no indication of any sizable lattice discretization errors.

For completeness note that the errors associated with the extracted values for $\Lambda_{\overline{\rm MS}}$ in Figure~\ref{FIG954} are rather large.
The reason for this is that we have here included the errors associated with the lattice spacings and also with $r_0 / a$  (cf.\ Table~\ref{TAB077}). This is absolutely essential when comparing results obtained at different values of the lattice spacing.

Figure~\ref{FIG954} also shows continuum extrapolations assuming a dependence $\propto a^2$ for both $r_0\Lambda_{\overline{\rm MS}}$ and $\Lambda_{\overline{\rm MS}}$. Within errors the results of these extrapolations agree with the results obtained for our smallest lattice spacing ($\beta = 4.35$, $a\approx 0.042\,{\rm fm}$). Even though no clear indication of any systematic upward or downward tendency as a function of $a^2$ is visible in Figure~\ref{FIG954}, we are conservative in our error estimate and infer additional systematic errors of $\pm 8 \, \textrm{MeV}$ for $\Lambda_{\overline{\rm MS}}$ and $\pm 0.010$ for $r_0\Lambda_{\overline{\rm MS}}$ to be accounted in our final result. These estimates are obtained by taking the difference between the central values of $r_0\Lambda_{\overline{\rm MS}}$ and $\Lambda_{\overline{\rm MS}}$ at our smallest lattice spacing and the corresponding continuum extrapolations.


\subsubsection*{Finite volume effects and non-vanishing light quark mass corrections}

Finite volume effects were investigated in detail in Ref.~\cite{Jansen:2011vv}. In this analysis such effects were found to be negligible compared to the other errors discussed above.

Similarly, potential corrections on $\Lambda_{\overline{\rm MS}}$ due to non-vanishing light quark masses were examined in detail in \cite{Jansen:2011vv} by studying the variation of $\Lambda_{\overline{\rm MS}}$ for different pion masses in the range $m_\textrm{PS} \approx
325 \, \textrm{MeV} \ldots 517 \, \textrm{MeV}$ at fixed lattice spacing and spacetime volume.
In the quark mass region investigated, $\Lambda_{\overline{\rm MS}}$ was found to be constant within tiny statistical errors of $\approx\pm 1 \,\textrm{MeV}$.
Therefore, we do not expect the non-vanishing light quark masses on the lattice to induce any significant deviations from the zero quark mass limit, for which the perturbative expressions in Sec.\ \ref{sec:perturbation} were derived.
In other words, we consider the systematic error introduced by comparing a lattice computation featuring massive light quarks with a perturbative calculation in the zero quark mass limit negligible compared to the other uncertainties discussed above.

For the above reasons we do not take any potential errors arising from finite volume effects and due to non-vanishing light quark mass corrections into account when quoting our final result for $\Lambda_{\overline{\rm MS}}$ in this paper.


\subsection{\label{SEC790}Final results for $\Lambda_{\overline{\rm MS}}$}

In the following, we present our final results for $\Lambda_{\overline{\rm MS}}^{(n_f=2)}$. We quote these results both in units of $\textrm{MeV}$ and in units of $r_0$. Since there seem to be up to $\mathcal{O}(10\%)$ unresolved differences regarding scale setting between different lattice QCD collaborations \cite{Sommer:2014mea}, we prefer the scale setting-independent quantity $r_0\Lambda_{\overline{\rm MS}}$.

The determination is based on the lattice results at our finest lattice spacing, i.e.\ at $\beta = 4.35$, as explained in Sec.~\ref{SEC444}. The individual error contributions, which we combine by adding them in quadrature, are
\begin{itemize}
\item[(1)] the correlated systematic errors associated with the unknown contributions beyond ${\cal O}(\alpha_s^4)$ of the perturbative $Q\bar Q$ static potential and the input parameters of the fitting procedure (cf.\ Sec.~\ref{SEC444}),

\item[(2)] the statistical errors associated with the $\chi^2$ minimizing fits,

\item[(3)] the estimated lattice discretization errors (cf.\ Sec.~\ref{SEC330}), and

\item[(4)] the errors associated with the lattice spacing $a = 0.0420(17) \, \textrm{fm}$ and with $r_0 / a = 9.81(13)$.
\end{itemize}
%
%

Taking all these contributions into account, we finally obtain
\begin{equation}
\label{EQN844}r_0\Lambda_{\overline{\rm MS}}^{(n_f=2)}\ = 0.692(31) \, , \quad  \Lambda_{\overline{\rm MS}}^{(n_f=2)} = 331(21) \, \textrm{MeV}.
\end{equation}
For completeness, note that being less conservative with regard to the estimate of the uncertainty associated with the truncation of the perturbative expressions and just taking NNNLO fits into account (cf. also Sec.~\ref{SEC444}), would result in $r_0\Lambda_{\overline{\rm MS}}^{(n_f=2)}= 0.665(16)$ and $\Lambda_{\overline{\rm MS}}^{(n_f=2)} = 318(16) \, \textrm{MeV}$.

The final results~\eqref{EQN844} for $\Lambda_{\overline{\rm MS}}$ extracted from our momentum space analysis are more precise than those obtained in position space using the same lattice data: $r_0\Lambda_{\overline{\rm MS}}^{(n_f=2)}= 0.658(55)$ and $\Lambda_{\overline{\rm MS}}^{(n_f=2)} = 315(30) \, \textrm{MeV}$.

Let us emphasize again that the errors quoted here and in \Eqref{EQN844} also account for contributions which are not directly related to the extraction of $\Lambda_{\overline{\rm MS}}$, as e.g.\ the uncertainties of $a$ and of $r_0 / a$. 
Hence, with regard to a comparison of the results for $\Lambda_{\overline{\rm MS}}$ obtained from an analysis of the $Q\bar Q$ static potential in momentum space as opposed to a similar analysis in position space, we consider the comparison in Sec.~\ref{SEC444} -- not accounting for these additional uncertainties -- as most relevant and meaningful.

For easy reference, we collect all error contributions for our momentum space analysis in Table~\ref{TAB490}, where we also confront them to the analogous error contributions for a position space analysis \cite{Jansen:2011vv}.

\begin{table}[htb]
\begin{center}
\begin{tabular}{|c|c|c|c|c|c|}
\hline
 & & & \vspace{-0.40cm} \\
error source & momentum  space & position space & comments and remarks \\
 & & & \vspace{-0.40cm} \\
\hline
 & & & \vspace{-0.40cm} \\
\hline
 & & & \vspace{-0.40cm} \\
(1) correlated    & $13 \, \textrm{MeV}$ & $20 \ldots 26 \, \textrm{MeV}$ & NNLO and NNNLO \\
\cline{2-4}
 & & & \vspace{-0.40cm} \\
systematic errors & $\phantom{0}3 \, \textrm{MeV}$  & $13 \ldots 14 \, \textrm{MeV}$ & NNNLO only \\
\hline
 & & & \vspace{-0.40cm} \\
(2) statistical & $\approx 2 \ldots 4 \, \textrm{MeV}$ & $\approx 2 \, \textrm{MeV}$ & The statistical error of the lattice \\
errors          &                                                      &                             & static potential is propagated \\
 & & & through to $\Lambda_{\overline{\rm MS}}$ via jackknife. \\
\hline
 & & & \vspace{-0.40cm} \\
(3) lattice           & $\ll 8 \, \textrm{MeV}$ & $\ll 6 \, \textrm{MeV}$ & These values amount to rather  \\
discretization errors &                       &                       &  conservative upper bounds.\\
\hline
 & \multicolumn{2}{|c|}{} & \vspace{-0.40cm} \\
(4) errors associated & \multicolumn{2}{|c|}{$\approx 13 \, \textrm{MeV}$} & $\approx \Lambda_{\overline{\rm MS}} \times (\Delta a / a) \approx \Lambda_{\overline{\rm MS}} \times 0.04$ \\
with the lattice      & \multicolumn{2}{|c|}{} & \\
spacing               & \multicolumn{2}{|c|}{} & \vspace{-0.40cm} \\
 & \multicolumn{2}{|c|}{} & \\
\hline
\end{tabular}

\caption{\label{TAB490}Error contributions for a momentum space analysis (this work) confronted to those for a position space analysis (Ref.~\cite{Jansen:2011vv}).}

\end{center}
\end{table}


\newpage

\section{\label{sec:Concl}Conclusions}

We have determined $\Lambda_{\overline{\rm MS}}^{(n_f=2)}$ by fitting perturbative expressions for the $Q\bar Q$ static potential at the presently best know accuracy, i.e. up to ${\cal O}(\alpha_s^4)$ in momentum space, to lattice results obtained at a rather fine lattice spacing $a \approx 0.042 \, \textrm{fm}$.

In contrast to previous works in this direction (cf.\ e.g.\ \cite{Michael:1992nj,Gockeler:2005rv,Brambilla:2010pp,Leder:2010kz,Jansen:2011vv,Bazavov:2012ka,Fritzsch:2012wq}) we have employed a discrete Fourier transform to transform the lattice results for the $Q\bar Q$ static potential to momentum space. The extraction of $\Lambda_{\overline{\rm MS}}$ by fitting perturbative expressions of the static potential to lattice results has exclusively been performed in momentum space.

Resorting to a previous position space extraction of $\Lambda_{\overline{\rm MS}}^{(n_f=2)}$ based on the same lattice data \cite{Jansen:2011vv}, we could show that the momentum space analysis allows for a more precise determination of $\Lambda_{\overline{\rm MS}}$.
We could reduce the associated errors by a factor of $\approx 1.5 \ldots 4$, depending on the details of the fitting procedures and the estimate of the error associated with the truncation of the perturbative expressions used (cf.\ in particular Sec.~\ref{SEC444}).
This improvement can mainly be attributed to the nearly perfect stability of the results for $\Lambda_{\overline{\rm MS}}$ with respect to variations of the momentum fitting range (cf.\ Figure~\ref{FIG003}).
Such behavior has not been observed in position space, where comparably rather strong variations have been observed (cf.\ also Figures~3 and 4 of \cite{Jansen:2011vv}).

In units of the hadronic scale $r_0$ our final result for $\Lambda_{\overline{\rm MS}}$ reads
\begin{equation}
\label{eq:final_} r_0\Lambda_{\overline{\rm MS}}^{(n_f=2)} = 0.692(31) ,
\end{equation}
while in physical units it is given by
\begin{equation}
\label{eq:final} \Lambda_{\overline{\rm MS}}^{(n_f=2)}  =  331(21) \, \textrm{MeV} .
\end{equation}
The quoted uncertainties include all sources of systematic error: the neglect of higher orders in the perturbative expansion, the dependence of the fit results on the fitting range $p_\textrm{min} \ldots p_\textrm{max}$ and the parameter $M$, lattice discretization errors, finite volume effects and uncertainties due to nonvanishing quark masses on the lattice. The uncertainties are dominated by the variations of $\Lambda_{\overline{\rm MS}}$, when switching from NNLO to NNNLO, and in the case of \Eqref{eq:final} by the uncertainties associated with the lattice spacing.

We note that the results (\ref{eq:final_}) and (\ref{eq:final}) compare well with other determinations of $\Lambda_{\overline{\textrm{MS}}}^{(n_f=2)}$ from the literature, e.g.\ \cite{DellaMorte:2004bc,Leder:2010kz,Gockeler:2005rv,Sternbeck:2010xu,Blossier:2010ky,Kneur:2011vi,Kneur:2013coa,Cichy:2013eoa} (cf.\ also the more detailed discussion and graphical summary in \cite{Jansen:2011vv}).

In the future it would be interesting to adopt a similar momentum space analysis to lattice results for the $Q\bar Q$ static potential with $n_f = 0$, $n_f = 2+1$ and $n_f = 2+1+1$ dynamical quark flavors.
For all these cases we expect a momentum space analysis to benefit from the better convergence behavior of the perturbative static potential in momentum space as compared to position space (cf. also the detailed discussion in the introduction, Sec.~\ref{sec:intro}), which -- as demonstrated in the present work -- can reduce the error on the extracted values of $\Lambda_{\overline{\rm MS}}$ as compared to a position space analysis of the $Q\Bar Q$ static potential.


\newpage

\section*{Acknowledgments}

We would like to thank Karl Jansen for helpful discussions, and the European Twisted Mass Collaboration for the generation of gauge link configurations used in this paper.

M.W.\ acknowledges support by the Emmy Noether Programme of the DFG (German Research Foundation), grant WA 3000/1-1.

F.K. is grateful to M. \& M.~Tingu for inspiring comments, and thankful to Karl-Heinz ``Serv$\bar{\rm u}$s'' Dubansky for providing a clean environment and Gummib\"aren.

This work was supported in part by the Helmholtz International Center for FAIR within the framework of the LOEWE program launched by the State of Hesse.



\end{document}